\documentstyle[11pt,aaspp4]{article}

\slugcomment{accepted by ApJ August 2001}

\lefthead{Leggett et al.}
\righthead{}

\begin{document}
 
\title{Infrared Photometry of Late--M, L, and T Dwarfs}

\author{S.~K.\ Leggett\altaffilmark{\ref{UKIRT}},
David A.\ Golimowski\altaffilmark{\ref{JHU}},
Xiaohui Fan\altaffilmark{\ref{IAS}},
T.~R.\ Geballe\altaffilmark{\ref{Gemini}},
G.~R.\ Knapp\altaffilmark{\ref{Princeton}},\\
J. Brinkmann\altaffilmark{\ref{APO}},
Istv\'{a}n Csabai\altaffilmark{\ref{JHU},\ref{Eotvos}},
James~E.\ Gunn\altaffilmark{\ref{Princeton}},
Suzanne L.\ Hawley\altaffilmark{\ref{UW}},
Todd J.\ Henry\altaffilmark{\ref{JHU},\ref{GSU}},\\
Robert Hindsley\altaffilmark{\ref{NRL}},
\v{Z}eljko Ivezi\'{c}\altaffilmark{\ref{Princeton}},
Robert H.\ Lupton\altaffilmark{\ref{Princeton}},
Jeffrey R. Pier\altaffilmark{\ref{USNOFS}},
Donald P. Schneider\altaffilmark{\ref{PennState}},\\
J.\ Allyn Smith\altaffilmark{\ref{Wyoming}},
Michael A.\ Strauss\altaffilmark{\ref{Princeton}},
Alan Uomoto\altaffilmark{\ref{JHU}},
D.~G.\ York\altaffilmark{\ref{Chicago}}
}

\newcounter{address}
\addtocounter{address}{1}
\altaffiltext{\theaddress}{United Kingdom Infrared Telescope, Joint Astronomy
Centre, 660 North A'ohoku Place, Hilo, Hawaii 96720 \\
email for Leggett: skl@jach.hawaii.edu
\label{UKIRT}}
\addtocounter{address}{1}
\altaffiltext{\theaddress}{
Department of Physics and Astronomy, The Johns Hopkins University,
   3400 North Charles Street, Baltimore, MD 21218 \\
email for Golimowski: dag@pha.jhu.edu
\label{JHU}}
\addtocounter{address}{1}
\altaffiltext{\theaddress}{Institute for Advanced Study, Einstein Drive, 
Princeton, NJ 08540 \\
email for Fan: fan@ias.edu
\label{IAS}}
\addtocounter{address}{1}
\altaffiltext{\theaddress}{Gemini Observatory, 670 North A'ohoku Place,
Hilo, HI 96720 \\
email for Geballe: tgeballe@gemini.edu
\label{Gemini}}
\addtocounter{address}{1}
\altaffiltext{\theaddress}{Princeton University Observatory, Princeton, NJ 08544 \\
email for Knapp: gk@astro.Princeton.edu
\label{Princeton}}
\addtocounter{address}{1}
\altaffiltext{\theaddress}{Apache Point Observatory, 2001 Apache Point Road, 
P.O.~Box 59, Sunspot, NM 88349
\label{APO}}
\addtocounter{address}{1}
\altaffiltext{\theaddress}{Department of Physics and Complex Systems, 
E\"{o}tv\"{o}s
University, P\'{a}zm\'{a}ny P\'{e}ter s\'{e}t\'{a}ny 1/A, Budapest, H-1117, 
Hungary
\label{Eotvos}}
\addtocounter{address}{1}
\altaffiltext{\theaddress}{Department of Astronomy, University of Washington,
Box 351580, Seattle, WA 98195
\label{UW}}
\addtocounter{address}{1}
\altaffiltext{\theaddress}{Department of Physics and Astronomy, Georgia State 
University,
Atlanta, GA  30303
\label{GSU}}
\addtocounter{address}{1}
\altaffiltext{\theaddress}{Remote Sensing Division,
United States Naval Research Laboratory, 
Washington, DC 20375
\label{NRL}}
\addtocounter{address}{1}
\altaffiltext{\theaddress}{Department of Physics and Astronomy,
University of Wyoming, 
P.O.\ Box 3905, Laramie, WY 82071
\label{Wyoming}}
\addtocounter{address}{1}
\altaffiltext{\theaddress}{Department of Astronomy and Astrophysics,
The Pennsylvania State University,
University Park, PA 16802
\label{PennState}}
\addtocounter{address}{1}
\altaffiltext{\theaddress}{University of Chicago, Astronomy \& Astrophysics
Center, 5640 S. Ellis Ave., Chicago, IL 60637
\label{Chicago}}
\addtocounter{address}{1}
\altaffiltext{\theaddress}{US Naval Observatory, Flagstaff Station, 
P.~O. Box 1149, Flagstaff, AZ 86002-1149
\label{USNOFS}}

\begin{abstract} 

We present $ZJHKL^{\prime}M^{\prime}$ photometry of a sample of 58 late--M, L, and T~dwarfs,
most of which are identified from the Sloan Digital Sky Survey and the Two Micron All--Sky
Survey.  Near--infrared spectra and spectral classifications for most of this sample are
presented in a companion paper by Geballe et al.  We derive the luminosities of 18 dwarfs in
the sample with known parallaxes, and the results imply that the effective temperature range
for the L~dwarfs in our sample is approximately 2200---1300~K and for the T~dwarfs 
1300---800~K.  We obtained new photometric data at the United Kingdom Infrared Telescope for:
42 dwarfs at $Z$, 34 dwarfs at $JHK$, 21 dwarfs at $L^{\prime}$, as well as $M^{\prime}$
data for two L dwarfs and two T dwarfs.  The $M^{\prime}$ data provide the first accurate
photometry for L and T~dwarfs in this bandpass --- for a T2 and a T5 dwarf, we find
$K$--$M^{\prime} = 1.2$ and 1.6, respectively.  These colors are much bluer than predicted by
published models suggesting that CO may be more abundant in these objects than expected, as
has been found for the T6 dwarf Gl~229B.  We also find that $K$--$L^{\prime}$ increases
monotonically through most of the M, L, and T subclasses, but it is approximately constant 
between types
L6 and T5, restricting its usefulness as a temperature indicator.  The degeneracy is probably
due to the onset of CH$_4$ absorption at the blue edge of the $L^{\prime}$ bandpass.  The
$JHK$ colors of L dwarfs show significant scatter, suggesting that the fluxes in these
bandpasses are sensitive to variations in photospheric dust properties.  The $H-K$ colors of
the later T dwarfs also show some scatter which we suggest is due to 
variations in pressure--induced H$_2$ 
opacity, which is sensitive to gravity and metallicity.

\end{abstract}
 
\section{Introduction}

For most of the twentieth century, the classical Harvard spectral types --- OBAFGKM ---
spanned the temperature range of all known main--sequence dwarfs.  The discoveries of
very cool companions to the white dwarf GD~165 (\cite{bz88}) and the M dwarf Gliese~229
(\cite{nak95}), however, foreshadowed the first extension of the Harvard system in
nearly 100 years.  GD~165B is now known to have $T_{\rm eff} \approx 1900$~K and a
probable mass of $\sim~0.07~M_{\odot}$, placing it near the stellar--substellar mass
boundary (\cite{k99b}).  With $T_{\rm eff} \approx 950$~K and a mass of
0.015---0.07~$M_{\odot}$ (\cite{sa00}), Gl~229B is cool enough to show methane
absorption in its near--infrared spectrum and is unambiguously substellar (i.e.,
a brown dwarf).  These two objects remained unique until the late 1990's, when new sky
surveys began revealing significant numbers of similar objects in the field.

In 1997, several GD~165B--like objects were discovered from two near--infrared sky surveys,
the Two Micron All--Sky Survey (2MASS; \cite{k97}) and the  DEep Near--Infrared Survey 
(DENIS; \cite{d97}), as well as a southern hemisphere proper motion survey (\cite{rla97}).  
In 1999, several Gl~229B--like brown dwarfs were found in the
commissioning data of the Sloan Digital Sky Survey (SDSS; \cite{str99}; \cite{tsv00}),
the 2MASS database (\cite{bur99}), and the New Technology Telescope Deep Field
(\cite{cub99}).  GD~165B and Gl~229B became the respective prototypes of
two new spectral classes, L and T (\cite{k99a}; \cite{mar99b}).  Over 100 L~dwarfs and
tens of T~dwarfs are now known.

In this paper, we present 0.95---4.70~$\mu$m photometry of 58 late--M, L, and T dwarfs
obtained with the United Kingdom Infrared Telescope (UKIRT).  We describe the sample in \S2
and present the photometry in \S3.  In \S4, we derive the luminosity and estimate the
temperature of 18 M, L and T dwarfs for which parallaxes are known, to place our sample in a
physical context.  In \S5 we discuss the correlation of the photometry with the spectral
sequence presented in a companion paper by Geballe et al.\ 2002 (hereafter G02); we also
discuss the correlations between colors, and we compare the $K-L^{\prime}$ and $K-M^{\prime}$
colors of the L and T dwarfs with model predictions.  Conclusions are given in \S6.

\section{The Sample}

Most of the sample comprises L and T dwarfs identified from SDSS (\cite{sdss}) and
2MASS (\cite{2m}).  G02 describe the selection criteria for candidate L and T dwarfs
culled from the SDSS photometric catalog.  The sample presented here consists of both
new and previously reported SDSS dwarfs, as well as other published cool dwarfs.  This
work is primarily concerned with the infrared colors of L and T~dwarfs, but
late--M~dwarfs are included to define a blue edge to the sequences under study.

Table 1 lists the 58 dwarfs in the sample.  Names and coordinates are given in
columns~1, 2, and 3.  Throughout this paper, we abbreviate the names of the DENIS,
2MASS, and SDSS objects by giving the survey acronym followed by the first four digits
of right ascension and the sign and first two digits of declination.  Note that
Gl~570D is also known as 2MASSW~J1457150--212148 and 2MASSW J1523+30 is also known as
Gl~584C.  Column~4 of Table 1 lists the distance moduli, where available, derived from
sources as given in the Table.  The spectral types are listed in column~5.

Of the twelve M dwarfs in the sample, eleven have previously published spectral types.
The one new SDSS M dwarf is classified as M8.5 by G02 on the basis
of near--infrared water band indices.  There are 29 L dwarfs in
the sample.  One is the prototype GD~165B (\cite{bz88}), another is Kelu-1
(\cite{rla97}), three are from DENIS (\cite{d97}), thirteen are from 2MASS
(\cite{k00}), and eleven are from SDSS (\cite{fan00}; G02).  G02 have classified all
but four of these L dwarfs based on near--infrared spectral indices; the classification
scheme is consistent with the optical/red schemes proposed by \cite{k99a} and \cite{mar99b},
at least for spectral types L0 to around L6 (see discussion in G02). 
The spectral types of the four L dwarfs not classified by us have been taken from
\cite{k00}.  There are 17 T~dwarfs in the sample. One is the prototype Gl~229B
(\cite{nak95}), eleven are from SDSS, and five from 2MASS.  G02 have classified all of
these objects using  near---infrared spectra.

\section{Photometric Data}

\subsection{Instrumentation}

All new photometric data were obtained at the 3.8~m UKIRT on Mauna Kea, over the period 1999
June to 2001 May.  Two cameras were used:  UFTI, which has filters covering the region
0.85---2.4~$\mu$m and a plate scale of 0$\farcs$091 pixel$^{-1}$; and IRCAM, which has
filters covering the region 1.15---4.9~$\mu$m and a plate scale of 0$\farcs$081 pixel$^{-1}$.
The IRCAM field of view is 20$\farcs$7; UFTI was used in single--quadrant readout mode resulting
in a field of view of 46$\farcs$6. 
UFTI contains a non--standard $Z$ filter.  Both cameras contain Mauna Kea Observatory Near
Infrared (MKO--NIR) $J$, $H$, and $K$ filters.  IRCAM also contains MKO--NIR $L^{\prime}$ and
$M^{\prime}$ filters.  The MKO--NIR system matches the near--infrared atmospheric windows,
thereby producing more accurate photometry than previous filter systems and easing comparison
of photometry between observatories (\cite{si01}; \cite{to01}).  The half--power wavelength
ranges of the filters are given in Table 2.  These bandwidths reflect the transmission
profiles of the filters at cold (instrument) temperatures convolved with the telescope optics
and the Mauna Kea atmospheric transmission; the response of the detectors is flat across each
filter bandpass and has not been included.  The filter profiles are shown in Figure 1.  To
compare these profiles with those of other commonly used filters, see for example \cite{bb88}.

The filters differ from the previous ``UKIRT''
system filters; transformations between the UKIRT $JHK$ system and the MKO--NIR system for
stars as cool as L dwarfs are given by \cite{haw01}.  These transformations cannot be applied
to the colors of T dwarfs because of their very structured flux distributions within the
filter bandpasses, and T dwarf  colors must be transformed by convolving the flux distributions
with the filter profiles, as described in \S 3.3.  The filter profiles and calibrated flux 
distributions for M, L, and T dwarfs can be obtained from the authors upon request.  The 
photometric system used in this paper is calibrated by adopting zero magnitude at all wavebands 
for Vega; an infrared spectrum of Vega is also available on request. 
Note that different photometric systems produce significantly different infrared magnitudes: 
at $JHK$ for L dwarfs the differences between for example the MKO--NIR and UKIRT, 
or MKO--NIR and 2MASS, magnitudes, are $\sim$5\%; for T dwarfs the difference at K is  
$\sim$10\%, at $H$ $\sim$5\% and at $J$ $\sim$30\%.  At $L^{\prime}$ and $M$, T dwarf magnitudes 
can vary by $\sim$20\% depending on the filter set used; this is discussed further in \S 5.3.

\subsection{Observations}

Table~3 lists the UKIRT $Z$ magnitudes and dates of observation for 42 dwarfs in the
sample.  To calibrate the non--standard $Z$ photometry, we derived transformations
between the SDSS $z'$, Cousins $I$, UKIRT $J$ and $Z$ filters using A0 stars, which
are defined as having zero color, and provisional $z'$ standards (\cite{kr98}).  In
doing so, we transformed the $z'$ magnitudes from the SDSS AB photometric system to a
colorless Vega--based system ($z^\dag$) using the equation
 $$ z^\dag = z' - 0.572 $$
(see Table~8 of \cite{fuk96}).  The derived transformations are:  
$$ Z_{\rm UFTI} = z^\dag - (0.34 \pm 0.03)(I_{\rm C}-z^\dag) $$ 
$$ Z_{\rm UFTI} = z^\dag - (0.21 \pm 0.03)(z^\dag - J_{\rm UKIRT}) $$ 
The 5\% errors in the $Z$ magnitudes are usually
dominated by the calibration error.  The $Z$ filter provides color information where
the flux distribution is rapidly rising for L and T dwarfs, and it is useful for
calibrating both red spectra and the short--wavelength end of near--infrared spectra.
Observations typically consisted of 120--250~s exposures, repeated three to five times
with small telescope offsets between exposures.  Flat fields were created through the
night by median filtering these sets of dithered exposures.  (Flat fields were created
for all filter and instrument combinations using this technique.)

Table 4 lists the MKO--NIR $JHK$ magnitudes, errors, camera selection, and date of
observation for 34 dwarfs.  The data were calibrated using UKIRT faint standards
appropriately transformed onto the MKO--NIR system (\cite{haw01}).  No color terms
could be detected between the cameras.  Observations typically consisted of five 60~s
exposures, with small telescope offsets between each exposure.

Table 5 lists the magnitudes and dates of observation for 21 dwarfs observed at $L^{\prime}$
and four dwarfs (two L, two T) observed at $M^{\prime}$.  All data were obtained using IRCAM
and calibrated using UKIRT bright standards, for which no transformations are required
between the UKIRT and MKO $L^{\prime}$ and $M^{\prime}$ systems.  The $M^{\prime}$ data were
obtained over two consecutive dry nights.  At these wavelengths the sky background is high
and exposure times have to be short.  Typically, each $L^{\prime}$ exposure consisted of 100
coadded 0.2~s integrations, and each $M^{\prime}$ exposure comprised 75 coadds of 0.12~s
integrations.  The telescope was offset slightly between exposures, adjacent pairs of frames
were subtracted to remove the rapidly varying and high background, and every four pairs of
differenced images were combined and divided by a flat field.  This process was repeated
until sufficient signal to noise was achieved.  For targets with $L^{\prime} \sim$ 11---13,
total integration times of 1 to 30 min were required to achieve 5\% photometry.  For objects
with $M^{\prime} \sim$ 12, about 1.3 hr of integration were required for 10\% photometry.
(Overhead is about a factor of two at these wavelengths.)  These data are the first accurate
$M$--band measurements of L and T dwarfs.  The only previously published $M$--band
measurement is for Gl 229B, for which \cite{mat96} obtained 7$\pm$3~mJy, or 10.9$\pm
\sim$0.5~mag.

\cite{bai01} have reported variability at the 5\% level in the I--band for M and L dwarfs, 
which they suggest is due to dust formation.  We have repeat observations for very few of our 
targets.  Three dwarfs have two sets of $JHK$ measurements: the L dwarfs SDSS 0107+00, SDSS0830+48 
and SDSS 1326--00.  Observations of the first two objects repeated to better than 3\% and the 
last repeated well at J,
but at H and K the values differed by 9\% which may not be significant.

The image quality during our runs was characterized by seeing of FWHM 0$\farcs$4---1$\farcs$0.
No new, close (within a few arcseconds), candidate companions were resolved for any of our 
targets, to a 2$\sigma$ detection limit of H$\approx$20.5.  
Distant objects in the fields have not yet been followed up.

\subsection{Synthesized Photometry}

For some of the dwarfs listed in Table~1 we have synthesized MKO--NIR photometry by
transforming data obtained with the previous UKIRT filter system or by convolving flux
calibrated spectra with the MKO--NIR bandpass profiles.  We derived the MKO--NIR $JHK$
magnitudes of three M4---M5 dwarfs (LHS~11, LHS~315, and LHS~333AB) from their respective
UKIRT system magnitudes using the transformations of \cite{haw01}.  We synthesized MKO--NIR
$JHK$ magnitudes for several late--M, L, and T dwarfs using the bandpass profiles and flux
calibrated spectra from Leggett et al. 2000a, 2001, and G02.  We calibrated the synthetic
magnitudes by convolving the energy distribution of Vega with the bandpass profiles and by
assigning zero magnitude to Vega in each bandpass.  The objects listed in Table~1 for which
synthetic $JHK$ magnitudes were derived are:  the M dwarfs LHS~36, LHS~292, LHS~3003,
TVLM~513--46546, and BRI~0021--0214; the L dwarfs 2MASP~J0345+25, Kelu-1, DENIS-P~J1058--15,
GD~165B, and DENIS-P~J1228--15AB; and the T dwarfs SDSS~1346--00 and SDSS 1624+00.  For the T
dwarf Gl~229B we synthesized all of the $ZJHKL^{\prime}$ magnitudes using the flux calibrated
spectrum of \cite{l99} (a recalibrated combination of the spectra of \cite{g96} and
\cite{opp98}).

\subsection{Observed Colors}

Table 6 lists all the new and transformed $ZJHKL^{\prime}M^{\prime}$ colors for the sample,
sorted by spectral type.   Absolute $K$ magnitudes are given for objects with
published parallax measurements.

Table 6 also includes the previously published MKO--NIR colors of SDSS~0539--00,
SDSS~0837--00, SDSS~1021--03, SDSS~1254--01 (\cite{l00b}), and Gl~570D (\cite{g01b}).  $JHK$
photometry is listed for three UKIRT M--dwarf photometric standards (LHS~2347, LHS~2397a, and
LHS~2924), obtained as part of a UKIRT program to observe standards with the MKO filter set.
Additional $L^{\prime}$ data for M dwarfs have been taken from \cite{l92} and \cite{l98}.
Additional $L^{\prime}$ data for L dwarfs have been taken from \cite{jon96} for GD~165B, from
\cite{l98} for Kelu-1 and from \cite{l01} for 2MASP J0345+25, DENIS-P J0205--11AB, DENIS-P
J1058--15 and DENIS-P J1228--15AB.  Some of these data were obtained with the UKIRT
$L^{\prime}$ filter which cuts on slightly redder than the MKO--NIR filter.  Although this is
significant for late L dwarfs and T dwarfs, as we discuss below, for these earlier type
objects the effect is less than the uncertainties in the published magnitudes.

\section{Luminosities and Effective Temperatures}

It is useful to estimate effective temperatures for our sample in order to interpret 
the trends seen in our observed colors.  To do so we use the observed luminosities,
as the theoretical determinations of radius and luminosity for low--mass stars and brown 
dwarfs are reasonably robust and independent of the atmospheric models.  $T_{\rm eff}$
can therefore be derived from integrated luminosity with reasonable assumptions for
radius.
 
Figure 2 shows $M_K$ versus $J$--$K$ (left panel) and $K$--$L^{\prime}$ (right panel)
for those dwarfs in our sample with published parallaxes.  In the left panel, we have
labeled the faintest L and T dwarfs; in the right panel, we have labeled the known
binaries and Kelu-1, which is superluminous but for which {\it Hubble Space Telescope}
observations show no resolved companion (\cite{mar99a}).  DENIS-P~J1228--15AB and
DENIS-P~J0205--11AB are composed of identical pairs of L dwarfs (\cite{mar99a};
\cite{l01}).  2MASSI~J0746+20AB consists of a pair of similar early L dwarfs, and
2MASSs~J0850+10AB consists of an L dwarf and probably a T dwarf (\cite{re01b}).  

$J$--$K$ reddens at first with decreasing $M_K$, or decreasing effective temperature, then
becomes bluer as the $K$--band CH$_4$ and H$_2$O absorptions become stronger (see for example
the spectra presented in G02).  $K-L^{\prime}$ increases approximately monotonically in this
figure and spectral types are indicated along the top of the right panel.  As we show later,
the near--infrared colors of L and T dwarfs do show some scatter with spectral type and more
trigonometric parallaxes must be obtained before we can adequately determine how these
variations affect the infrared color--magnitude diagrams of brown dwarfs.

We have compiled the bolometric luminosities of 18 dwarfs in our sample for which both
flux--calibrated spectra and parallaxes are available.  The luminosities of the M to
mid--L dwarfs are obtained from Leggett et al. 2000a, 2001; those of the T dwarfs
Gl~229B and Gl~570D are from \cite{sa00} and \cite{g01b}, respectively.  We have
determined the luminosities of 2MASSW~J0036+18 (L4), 2MASSI~J0825+21 (L6),
2MASSW~J1523+30 (L8)  and 2MASSW~J1632+19 (L7.5), by summing their energy distributions
from the red to the $K$ band, interpolating the flux between the $K$ band and the
effective $L^{\prime}$ flux computed from our photometry, and assuming Rayleigh--Jeans
curves longward of $L^{\prime}$.  Neither the interpolation between $K$ and
$L^{\prime}$ nor the Rayleigh--Jeans extrapolation is a correct assumption for T
dwarfs, as there is CH$_4$ absorption shortward of $L^{\prime}$, and
there are CH$_4$ and H$_2$O absorption bands between 6 and 8~$\mu$m (\cite{bur97}).  A
correction to this simple approach was determined for Gl~229B by \cite{l99} using model
atmospheres.  \cite{l01} also used models to show that no correction is needed for
dwarfs as late as mid L.  Therefore, we adopt a correction for the 2MASS L7.5 and L8
dwarfs that is half that computed for Gl~229B, which amounts to a 5\% adjustment to the
total integrated luminosity.

Table 7 lists the names, spectral types, $K$--band bolometric corrections (BC$_K$), and
bolometric luminosities (expressed as log$_{10}~L/L_{\odot}$) for the 18 dwarfs.  The
uncertainties of the total measured fluxes are typically 5\%, and are usually dominated by
the absolute calibration.  These uncertainties correspond to errors in the computed BC$_K$
and log$_{10}~L/L_{\odot}$ of about 0.07 mag and 0.03 dex, respectively.  The parallaxes of
GD~165B and 2MASSI~J0825+21 are less well determined than those of the other dwarfs, so
their bolometric luminosities are correspondingly less certain.  Figure~3 shows BC$_K$
versus $J$--$K$ (left panel) and $K$--$L^{\prime}$ (right panel) so that $M_{bol}$ ($=$
BC$_K + M_K$) can be estimated for other dwarfs.  The sample is too small to allow a
polynomial fit to these data.

We used the relationship between $T_{\rm eff}$ and log$_{10}~$L/L$_{\odot}$ derived from
the models of \cite{cbah00} (see Figure~12 of \cite{l01}) to compute $T_{\rm eff}$ for
2MASSW~J0036+18, 2MASSI~J0825+21, 2MASSW~J1523+30  and 2MASSW~J1632+19, assuming their ages 
lie in the range 0.1---10~Gyr.  The computed values of $T_{\rm eff}$ for these dwarfs are 
listed in Table~7 along with previously published values for the other dwarfs (\cite{g01b};
Leggett et al. 2000a, 2001; \cite{sa00}).  The range in $T_{\rm eff}$ listed for each
dwarf reflects its uncertain age.  The temperatures derived for the late--L dwarfs
are consistent with the temperature at which CH$_4$ is expected to become more abundant
than CO in brown dwarf photospheres (\cite{lo99}); the appearance of  CH$_4$ absorption
features in both of the $H$ and $K$ bands is the defining spectral signature of a T dwarf
(G02).  A trend between spectral type and effective temperature is apparent in Table 7,
however as we show below, the $JHK$ colors of L dwarfs  show significant scatter 
with spectral type and more bolometric luminosities must be determined to confirm
whether or not a unique value of $T_{\rm eff}$ is associated with each spectral type.

\section{Discussion of Infrared Colors}

\subsection{Colors and Spectral Type}

Figure 4 is a plot of $J$--$H$ (top panel), $J$--$K$ (middle panel) and $K$--$L^{\prime}$
(bottom panel) versus spectral type for the dwarfs in our sample.  All the L and T dwarfs
have been classified spectroscopically by G02 except for four L dwarfs whose types have
been taken from \cite{k00}; these four objects are shown as open triangles in the figure. 
There is good consistency between L spectral types assigned using red spectra by Kirkpatrick 
et al. and those assigned using near--infrared spectra by Geballe et al.  for spectral
types L0 to around L6, however for the latest L types the schemes can differ by up to two
subclasses.
The average uncertainties are represented by the error bars shown in the left of each panel. 

At these wavelengths the principle opacity sources in the photospheres of cool dwarfs (with
$2700~\gtrsim~T_{\rm eff}~\gtrsim~800$~K) are:  H$_2$O at both the short and long wavelength
edges of each of the $JHK$ bands, CH$_4$ at 1.6--1.8, 2.2--2.4 and 3.1--3.6~$\mu$m
(affecting $H$, $K$ and the blue edge of the $L^{\prime}$ band), pressure--induced H$_2$ at
1.8--3.0~$\mu$m ($K$) and CO at 2.3--2.4 and 4.4--5.0~$\mu$m ($K$ and $M^{\prime}$).  Figure
15 in \cite{burrows01} shows absorption cross sections versus wavelength from the optical to
the $M$ band for all these species, and is a very useful reference.  The features can be
compared to our filter bandpasses shown in Figure 1.

T~dwarfs (with $1300~\gtrsim~T_{\rm eff}~\gtrsim~800$~K) become bluer in both $J$--$H$ and
$H$--$K$ as $T_{\rm eff}$ decreases and the $H$-- and $K$--band absorptions by CH$_4$
strengthen (G02).  The trend towards bluer $J$--$K$ begins around spectral type L8.  This
result is consistent with the observations of G02 that CH$_4$ absorption is seen in the
$K$--band spectra of the latest L~dwarfs.  Note that the blueward trend in $J$--$H$ slows
between types T6 and T8.  This behavior probably reflects the saturation of the CH$_4$
absorption in the $H$--band (see the spectra presented in G02).

$K$--$L^{\prime}$ generally increases through the M, L and T classes but the increase slows
for types around L6 through to T5.  \cite{noll00} detect the onset of 3.3~$\mu$m absorption
by CH$_4$ (the fundamental band) for 2MASSI~J0825+21 (L6) and 2MASSW J1507--16 (L5), and this
feature is strong for T dwarfs as shown by the spectrum presented by \cite{opp98}.  This
absorption band is included at the blue edge of our $L^{\prime}$ bandpass and so is probably
slowing the increase in $K$--$L^{\prime}$ for objects later than L6; tests with the
Oppenheimer et al.  $L$--band spectrum of Gl~229B show the effect to be
$\gtrsim$~0.2~mag at T6.  The more rapid rise in $K$--$L^{\prime}$ at later types 
probably reflects the saturation of the CH$_4$ absorption.

The colors are well constrained among the late M dwarfs.  However the $JHK$ colors of the L
dwarfs show significant scatter.  The same phenomenon is seen in Figure~5 of \cite{re01a}.
These spectral types have a range of effective temperature ($2200~\gtrsim~T_{\rm
eff}~\gtrsim~1300$~K) over which grain condensation is expected in the photosphere
(\cite{tsu96}; \cite{cbah00}; \cite{ack01}; \cite{all01}).  The brightness
temperature spectra presented by \cite{ack01} show that the near--infrared fluxes of L dwarfs
may be strongly affected by dust cloud behavior.  The absorbing dust is expected to heat the
photosphere and consequently reduce the depth of the H$_2$O absorption bands (for the
limiting cases of fully dusty and dust--free atmospheres compare Figures 8 and 14 in
\cite{all01}).  Table~6 shows that both $J$--$H$ and $H$--$K$ can vary by 0.2---0.3~mag among
L~dwarfs of the same type (repeat observations of two of the redder L dwarfs show good agreement,
\S 3.2, and so the reddening does not seem to be a temporary effect).  One example of an
apparently red L dwarf is the L5 SDSS~2249+00, for which G02 find  spectral 
types inferred separately from the $H$-- and $K$--band indices that are discrepant by three
subclasses, whereas these indices are consistent for the other L5 dwarfs.  However there are
also examples of L dwarfs with very different infrared colors but well--behaved spectral indices,
such as  the L3 dwarfs 2MASSI~J0028+15 and DENIS-P~J1058--15.  G02 find that the L
subclasses correlate well with the depth of the H$_2$O absorption band at 1.5~$\mu$m,
although the $K$--band water features shows less sensitivity to spectral type.
The scatter in
color with L subclass may indicate varying dust properties caused by differences in
metallicity (which affects dust abundance), age (which limits settling time), and rotational
velocity (which may inhibit dust settling).    A better
understanding of the spectroscopic and photometric behavior of mid--L dwarfs will have to
await detailed models with a full and accurate treatment of grain condensation.

The T dwarfs also show some scatter, in particular in $J$--$K$ and $K$--$L^{\prime}$ for the
latest T dwarfs.  At these temperatures ($T_{\rm eff}~\sim~1000$~K) grains are calculated to
lie below the photosphere, and the absorption bands of H$_2$O and CH$_4$ are close to saturation
(see G02).  The scatter seen here may be due to variations in the strength of
the pressure--induced H$_2$ opacity in the $K$--band, which is  expected to be an important 
opacity source at these temperatures and is sensitive to gravity
and metallicity (e.g. \cite{bor97}).  Again, understanding this behavior must await more 
detailed photospheric models.

\subsection{$ZJHKL^{\prime}$ Color--Color Sequences}

Figure 5 shows the $J$--$H$ versus $H$--$K$ diagram for our sample.  The average 
photometric errors are represented by the error bars
shown in the lower right corner of the figure.  The colors of the L and T dwarfs show 
some scatter as discussed in the previous section.  
Note that the early T~dwarfs are not easily distinguished from 
M and L~dwarfs by $JHK$ photometry alone. 

Figure 6 shows $J$--$K$ versus $Z$--$J$ (left panel) and $K$--$L^{\prime}$ (right panel) for
our sample.  While $Z$--$J$ correlates well with $J$--$K$ for M dwarfs, it is almost constant
($Z$--$J \approx 1.8$) for L and  T dwarfs.  This behavior is caused by the balanced
losses of flux in the $Z$ and $J$ bands from two very different sources of opacity.  As
$T_{\rm eff}$ decreases, increased absorption by the highly pressure--broadened K~{\small I}
resonance doublet at 0.7665 and $0.7699~\mu$m dominates the $Z$--band region of the spectrum
(\cite{tsu99}; \cite{bur00}; \cite{lie00}).  At the same time, deepening absorption by the
$1.15~\mu$m and, to a lesser extent, the short wavelength wing of the $1.4~\mu$m H$_2$O
bands, reduces the emergent $J$-band flux.  (See the spectra presented in G02 and the
bandpasses shown in Figure~1.)  The strong K~{\small I} feature makes the use of a red
continuum index problematical for classifying late L and T~dwarfs (G02).

Good correlation exists between $J$--$K$ and $K$--$L^{\prime}$ for types mid--M to
mid--L and for the later T subclasses.  The degeneracy of $K$--$L^{\prime}$ between types 
L6 and T5 can probably be attributed to increasing absorption in both bandpasses, as 
described above in \S 5.1.

\subsection{$KL^{\prime}M^{\prime}$: Discrepancy with Models}

In Table 8 we summarise our observed $K$--$L^{\prime}$ and $K$--$M^{\prime}$ colors as a
function of spectral type and estimated effective temperature.  We also compare the observed
colors to those calculated by two models which handle grain condensation (dust) differently.
These are the models of very low mass stars and brown dwarfs with dusty atmospheres by
\cite{cbah00}, and the models of brown dwarfs in which dust is treated as if condensed below
the photosphere by \cite{bur97}.  These structural models provide constraints on surface
gravity as a function of age and temperature, and the two models give consistent results
for age, temperature and gravity.  The range of age or gravity for the calculated colors 
are given at the bottom of the table.  

Table 8 shows that the $K$--$L^{\prime}$ colors for L dwarfs calculated by Chabrier et al.
agree reasonably well with our observations.  For T dwarfs their calculated $K$--$L^{\prime}$
colors are redder than observed while those of Burrows et al.  more closely match our
observations.  The $L^{\prime}$ bandpass used here has a bluer cut--on wavelength than that
used by either model  --- compare for example the bandpass shown in Figure 1 to that of
$L^{\prime}_{\rm AAO}$ in Figure 9 of \cite{bb88}.  Our bandpass will include some of the
CH$_4$ absorption seen in late L and T dwarfs (see also \S 5.1).  \cite{ds01} show that this
difference in bandpass can lead to the model $K$--$L^{\prime}$ colors being redder by 0.1~mag
for late L dwarfs and by 0.2~mag for mid--T dwarfs; we have confirmed the latter by tests
using the observed Gl~229B spectrum.  The large discrepancy between the observed $L^{\prime}$ fluxes
of T~dwarfs and those predicted from the models of Chabrier et al.  (1.3~mag at T6) is most
likely due to their treatment of dust; the authors acknowledge that dust grains are more
likely to exist below, rather than within, the photosphere of T dwarfs, as is assumed by
the Burrows et al.  models.

The observed $K$--$M^{\prime}$ colors of L and T dwarfs are bluer than calculated by either
model.  The absolute $K$ magnitudes predicted by the models are consistent with our
measurements, at least for the small sample of L and T dwarfs with known parallaxes.  Thus,
if the predicted $K$--$M$ colors are too red, then the predicted $M$--band fluxes are likely
to be too large.  For T dwarfs, the $K$--$M^{\prime}$ discrepancy of $\gtrsim$1.0~mag implies
that the models of both \cite{bur97} and \cite{cbah00} predict around three times the observed
$M$--band flux.  This is disappointing as model calculations had suggested that the 5$\mu$m 
region would be the preferred wavelength for brown dwarf studies using planned space instruments
(\cite{burrows01}).  The discrepancy found here implies that  achieving this goal will require
higher sensitivity than originally thought.

Differences in the $M$ filter passbands used for the model
calculations and the observations cannot explain the discrepancy.  The $M^{\prime}$
filter used here has half--power wavelengths of 4.57~$\mu$m and 4.80~$\mu$m (see Figure~1).
The typical $M$ filter has a half--power bandpass defined by the wavelength range
4.50---5.00~$\mu$m (see Figure 9 of \cite{bb88}).  No strong features are expected in the spectra 
of  the A0 calibration star or T~dwarfs within these wavelength ranges; for T dwarfs H$_2$O absorption 
will be important longward of the $M$ band  and CO absorption is not expected as all the carbon should
be in the form of CH$_4$ (but see the discussion below).
Allard (2001, private communication), using models similar to those in Chabrier et
al., finds no significant difference between $K$--$M$ and $K$--$M^{\prime}$.
Calculations we have done with model spectra made available to us by F. Allard and P. Hauschildt, and 
by D. Saumon and M. Marley, show the flux through the $M$ and $M^{\prime}$ filters differs only
by 4\% at $T_{\rm eff}=2000$~K and 9\% at $T_{\rm eff}=800$~K.  Tests using the observed
spectrum for the T dwarf Gl~229B imply the effect to be larger but still only 20\%.

The models of \cite{bur97} and \cite{cbah00} make very different assumptions about the presence
of dust in the photosphere, however, while they calculate  very different
$K$--$L^{\prime}$ colors at $T_{\rm eff}\approx 1000$~K, their $K$--$M^{\prime}$ colors
are very similar.  Hence it seems that dust does not cause the $M^{\prime}$ discrepancy observed here.
Another possible uncertainty in the models is treatment of molecular opacities.
\cite{noll97} and \cite{opp98} have detected CO absorption features in the $M$--band
spectrum of the T6 dwarf Gl~229B.  The implied abundance of CO in the photosphere of Gl~229B is much
greater than that expected under conditions of thermochemical equilibrium (\cite{fl96}).
This overabundance of CO may be the result of vertical mixing in the atmosphere, a condition
not included in the models of \cite{bur97} or \cite{cbah00}.  Opacity from CO, and/or the
change in the temperature---pressure structure of the atmosphere introduced by mixing, may
at least partly account for the $M$--band discrepancy between observations and models.
Convolving the $M^{\prime}$ filter profile with the Gl~229B spectrum of Noll et al.\ 1997
shows that the CO absorption bands reduce the $M'$ flux by $\sim$0.4~mag; convolving with
the wider $M$ filter shows the effect to  be $\sim$0.7~mag.  The latter is close to the observed
discrepancy if $T_{\rm eff}\approx 1300$~K (see Table 8).
Further study is clearly warranted.

\section{Conclusions}

We present $ZJHKL^{\prime}M^{\prime}$ photometry for a sample of 58 late--M, L, and
T~dwarfs, most of which have been classified spectroscopically by G02.  A well--defined 
but relatively new photometric system is used; differences between this and previous
near--infrared systems are significant  at around the 5\% level, but in the case of the 
T dwarfs the difference at $J$ is $\sim$30\% and at $L^{\prime}$ and $M/M^{\prime}$
it is $\sim$20\%.  Differences between photometric systems must be taken into account when comparing
infrared colors of L and T dwarfs taken from the literature.

We have determined
effective temperatures and bolometric corrections where possible, to show that
$2200~\gtrsim~T_{\rm eff}~\gtrsim~1300$~K for L dwarfs and $1300~\gtrsim~T_{\rm
eff}~\gtrsim~800$~K for the known T dwarfs.  The temperatures derived for late--L dwarfs are
consistent with the expected CO to CH$_4$ transition temperature in brown
dwarf atmospheres (\cite{lo99}).

We have analyzed the data by examining color--color, color--magnitude and color--spectral type
diagrams. We conclude the following, in order of increasing wavelength:

We find that a filter centered around $1~\mu$m (e.g., UKIRT $Z$) is well suited for
measuring the fluxes of brown dwarfs in a region where their energy distributions are
rapidly rising.  Consequently, the $Z$ filter is useful for flux calibrating red spectra, as
long as the bandpass is accurately known.  On the other hand, the pressure--broadened wings
of the K~{\small I} resonance doublet at $0.77~\mu$m and the increasing water absorption in
the $J$--band cause $Z$--$J$ to saturate at $\sim 1.8$ for L and T dwarfs.  Therefore,
$Z$--$J$ is not useful as a discriminator of brown dwarf spectral type.
  
The onset of CH$_4$ absorption at the L--T spectral boundary (see G02) causes the $J$--$H$ and
$H$--$K$ colors of T~dwarfs to become progressively bluer,  overlapping the
color space occupied by M dwarfs. The $ZJHK$--band
color trends of late--M, L, and T dwarfs show that, although SDSS and 2MASS broadband colors
together are superb for identifying {\it candidate} brown dwarfs, confirmation and
classification of all but the later T~dwarfs require follow--up spectroscopy or longer
baseline photometry.

We find significant scatter in the $JHK$ colors of both L and the later T dwarfs.  We suggest 
that the $JHK$ magnitudes of L dwarfs are sensitive to the presence
and composition of dust clouds in their photospheres.  We also suggest that the $K$ 
magnitudes of T dwarfs are sensitive to variations in gravity and/or metallicity
which affect the strength of the pressure--induced H$_2$ opacity in this band.
The observed scatter between color and spectral type may imply that there is not a unique 
relationship between effective temperature and spectral type; more bolometric luminosities are 
needed to determine whether this is true.

We find that $K$--$L^{\prime}$ increases monotonically through most of the M, L, and T
classes.  A color degeneracy of $K$--$L^{\prime} \approx 1.5$ exists between spectral types
L6 and T5, most likely due to CH$_4$ absorption at the blue edge of our $L^{\prime}$
bandpass, which has been shown by \cite{noll00} and \cite{opp98} to be detectable for
spectral types around L6 and later.   This limits the usefulness of $K$--$L^{\prime}$ as an
indicator of effective temperature.  Of all the color--color combinations spanned by our
near--infrared photometry, $J$--$K$ versus $K$--$L^{\prime}$ is the best discriminator of
late--M, L, and T spectral types.  Except for dwarfs near the L--T boundary, the dwarfs in
our sample can be classified photometrically in this color space within one or two
subclasses.

We have obtained accurate $M^{\prime}$ magnitudes for two L dwarfs and two T dwarfs.  The
$M$--band fluxes for T dwarfs deduced from our photometry are around three times fainter than
predicted by current atmospheric models.  This discrepancy may be caused in part by increased
opacity due to a higher than expected photospheric abundance of CO, as has been detected by
\cite{noll97} and \cite{opp98} for Gl~229B.  
This discrepancy demonstrates the need for improved models of substellar atmospheres
and for more  5~$\mu$m spectroscopic data, although the results presented 
here show that such observations will be more difficult than anticipated.

\acknowledgments We are very grateful to the staff at UKIRT for their assistance 
in obtaining the data presented in this paper.  Some 
data were obtained through the UKIRT Service Programme.  UKIRT is operated by 
the Joint Astronomy Centre on behalf of the U.~K.\ Particle
Physics and Astronomy Research Council.  
We are grateful to the referee for helpful comments on the manuscript, to
France Allard for calculations of colors through the M$^{\prime}$ bandpass,
to Denise Stephens for prepublication results, and to France Allard, Peter
Hauschildt, Didier Saumon and Mark Marley for making models available for
investigating filter bandpass effects.
XF acknowledges support from NSF grant PHY-0070928 and a Frank and Peggy
Taplin Fellowship.  GRK is grateful for support to Princeton University and
the Humanities Council, and to NASA via grants NAG5-6734 and NAG5-8083.
The Sloan Digital Sky Survey (SDSS) is a joint project of The
University of Chicago, Fermilab, the Institute for Advanced Study, the
Japan Participation Group, The Johns Hopkins University, the
Max-Planck-Institute for Astronomy (MPIA), the Max-Planck-Institute
for Astrophysics (MPA), New Mexico State University, Princeton
University, the United States Naval Observatory, and the University of
Washington. Apache Point Observatory, site of the SDSS telescopes, is
operated by the Astrophysical Research Consortium (ARC).
Funding for the project has been provided by the Alfred P. Sloan
Foundation, the SDSS member institutions, the National Aeronautics and
Space Administration, the National Science Foundation, the
U.S. Department of Energy, the Japanese Monbukagakusho, and the Max
Planck Society. The SDSS Web site is  {\tt http://www.sdss.org/}.

\newpage 

\begin{figure}
\epsscale{.7}
\plotone{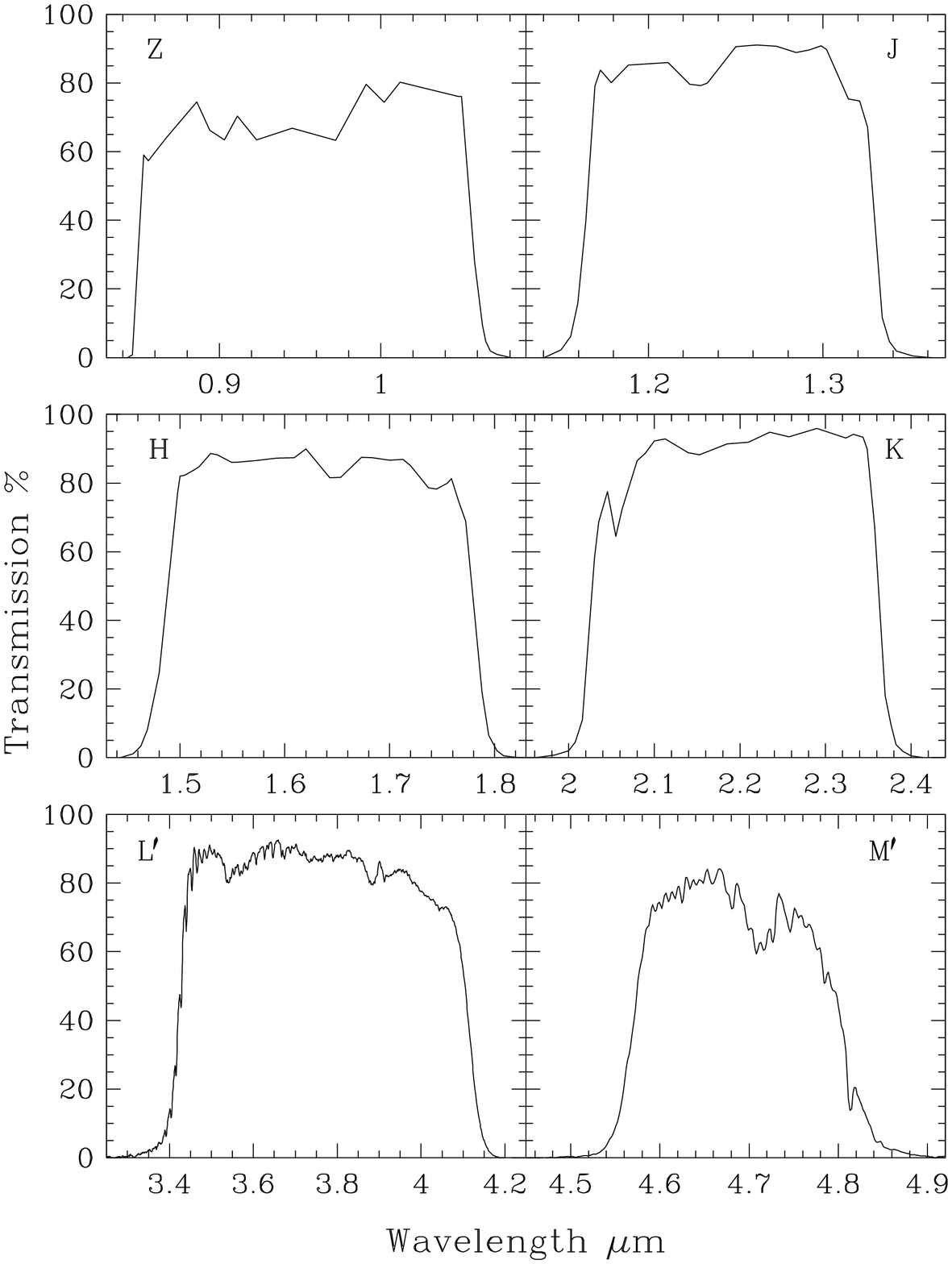}
\caption{Profiles of filters at camera temperature and convolved with 
atmospheric and telescope transmission.
\label{fig1}}
\end{figure}
\newpage

\begin{figure}
\plotfiddle{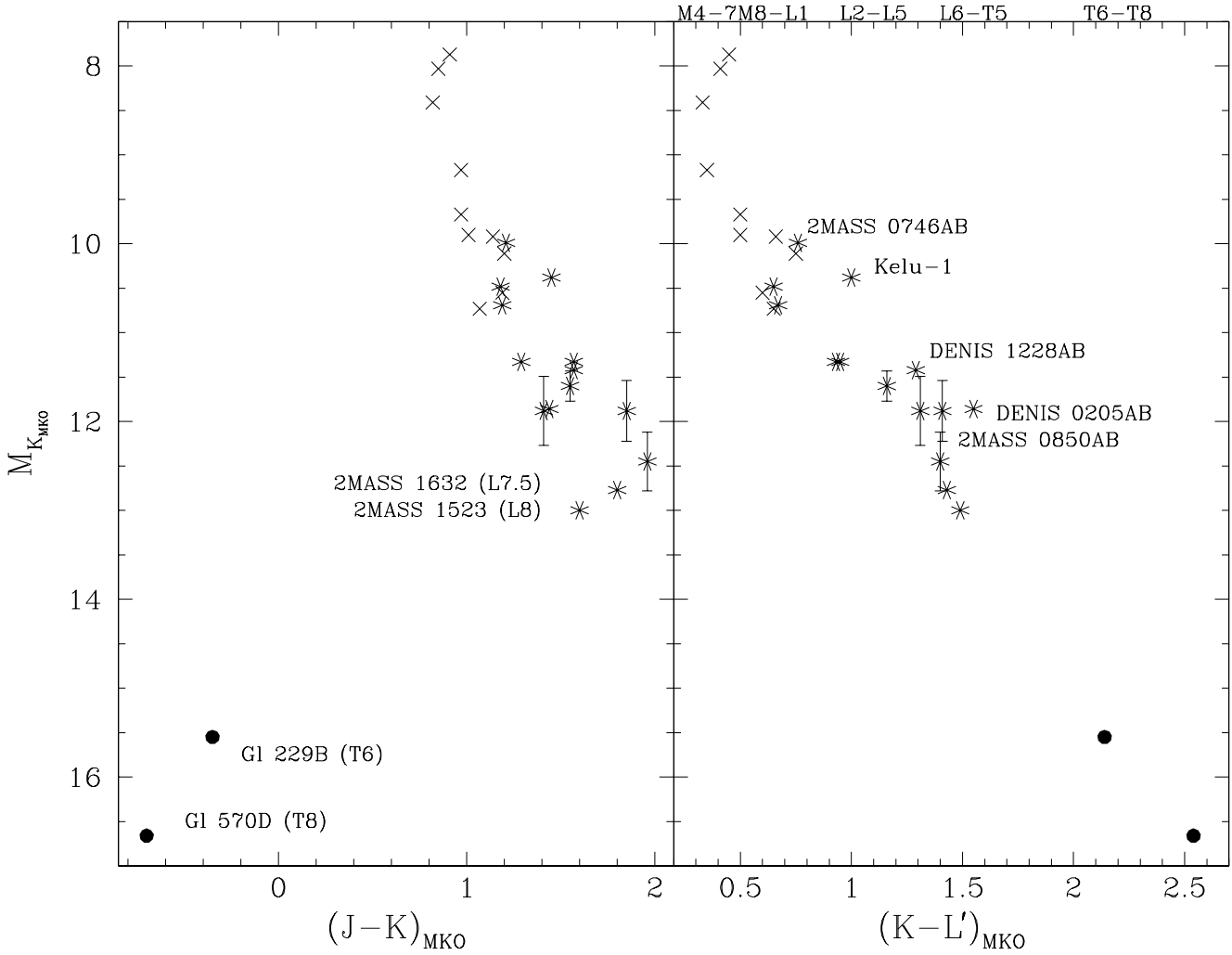}{14truecm}{-90}{70}{70}{-300}{450}
\caption{ Plots of absolute $K$ magnitude versus $J$--$K$ (left) and 
$K$--$L^{\prime}$ (right) for M~dwarfs (crosses), L~dwarfs (asterisks),
and T~dwarfs (filled circles).  Error bars are shown for dwarfs with absolute magnitude
errors $>0.15~$mag.
\label{fig2}}
\end{figure}
\newpage

\begin{figure}
\plotfiddle{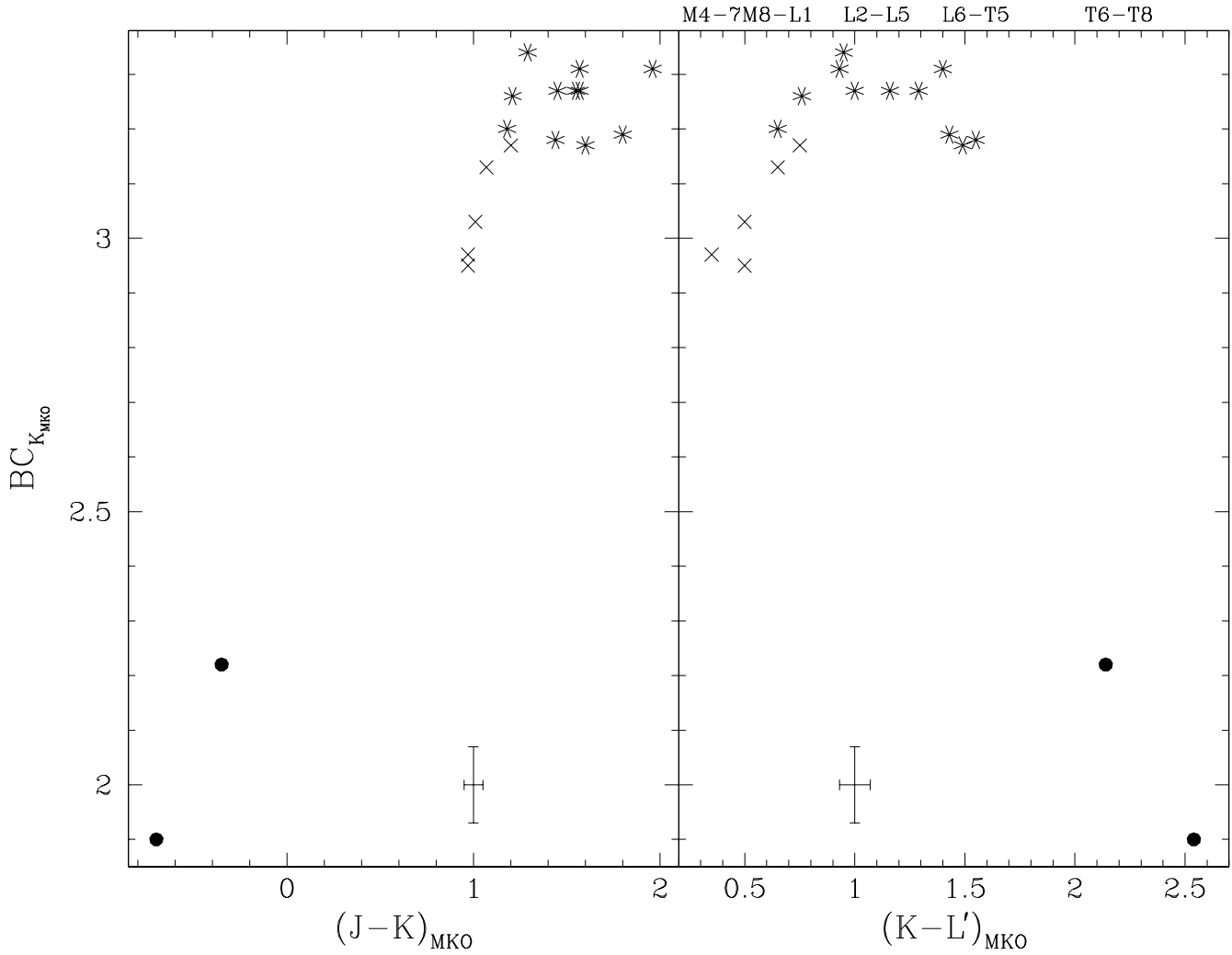}{14truecm}{-90}{70}{70}{-300}{450}
\caption{Plots of $K$-band bolometric correction versus $J$--$K$ (left) and 
$K$--$L^{\prime}$ (right) for M~dwarfs (crosses), L~dwarfs 
(asterisks), and T~dwarfs (filled circles).  A typical error bar is shown at the
bottom of each panel.
\label{fig3}}
\end{figure}

\newpage

\begin{figure}
\plotfiddle{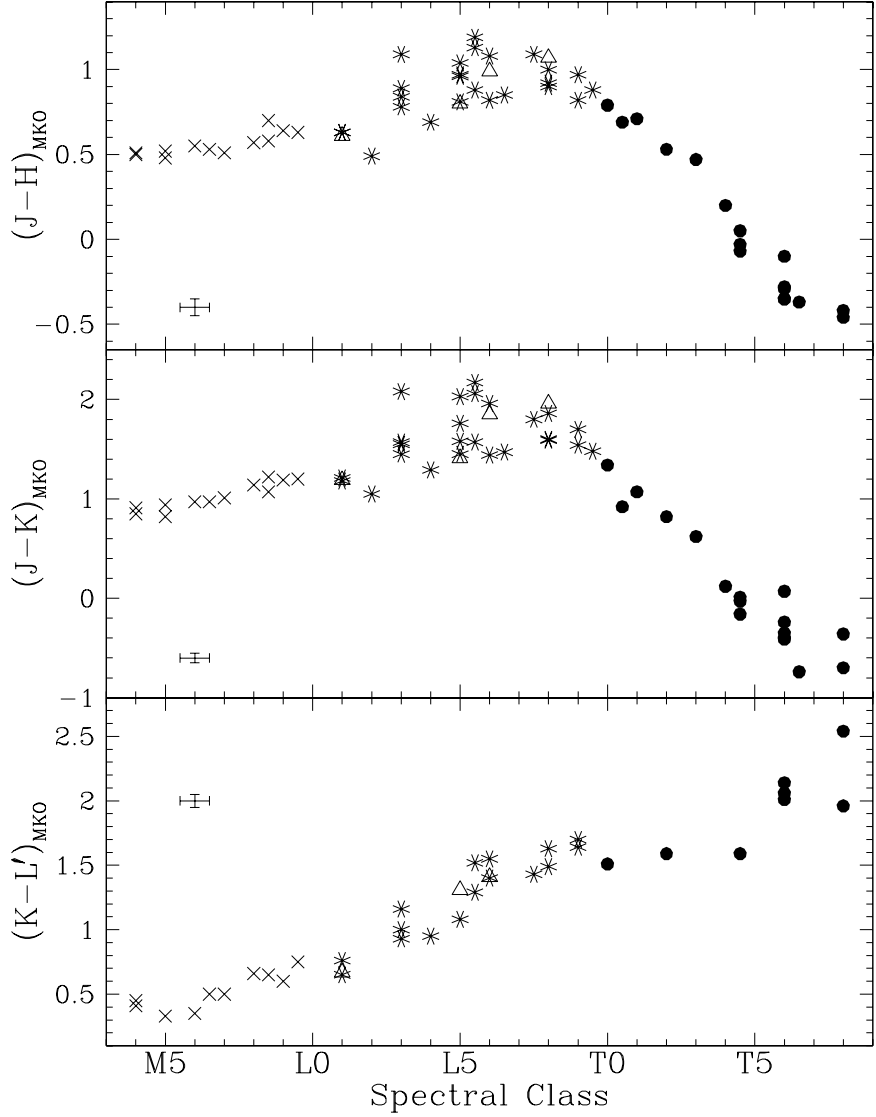}{14truecm}{0}{60}{60}{-200}{0}
\caption{ Plot of  $J$--$H$ (top),  $J$--$K$  (middle) and $K$--$L^{\prime}$ (bottom)
versus spectral type for M~dwarfs (crosses), L~dwarfs (asterisks), and T~dwarfs 
(filled circles) where spectral types are assigned by G02.  In addition
four L dwarfs are shown as open triangles for which spectral types are from \cite{k00}.
\label{fig4}}
\end{figure}

\newpage

\begin{figure}
\epsscale{.7}
\plotone{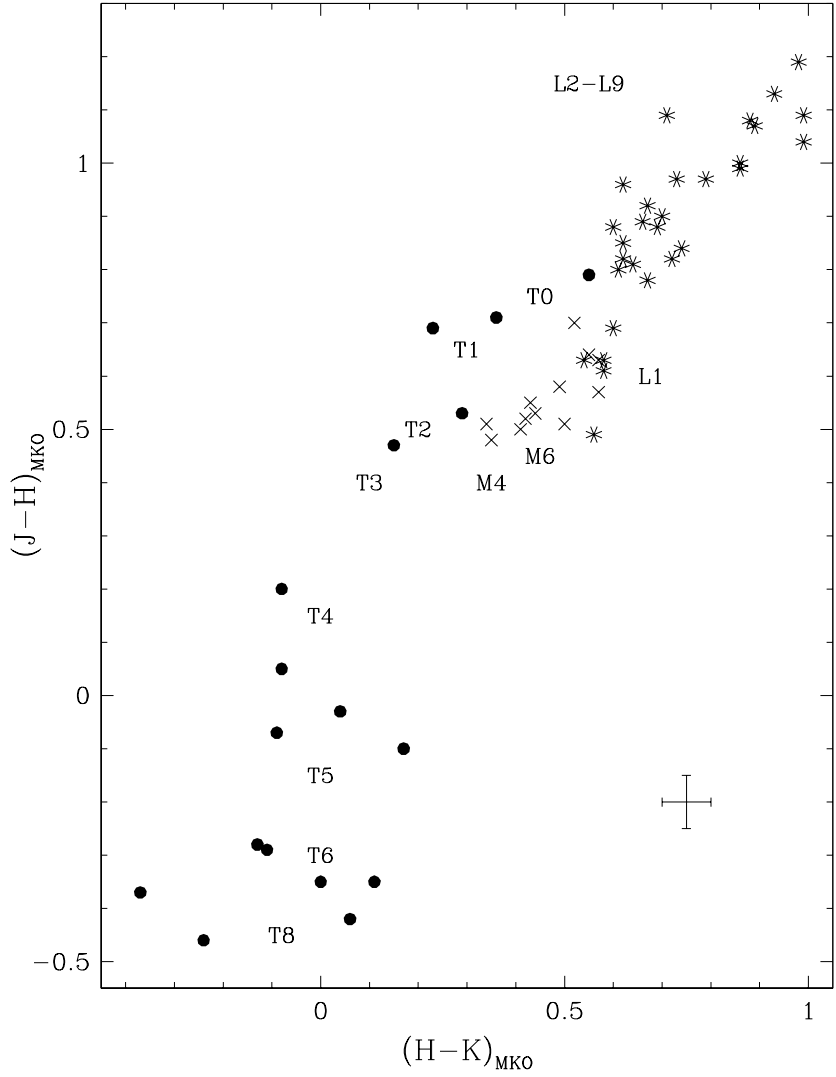}
\caption{ Color--color diagram of $J$--$H$ versus $H$--$K$ diagram for M~dwarfs 
(crosses), L~dwarfs (asterisks), and T~dwarfs (filled circles).  The location by
spectral type is indicated.
\label{fig5}}
\end{figure}

\newpage

\begin{figure}
\plotfiddle{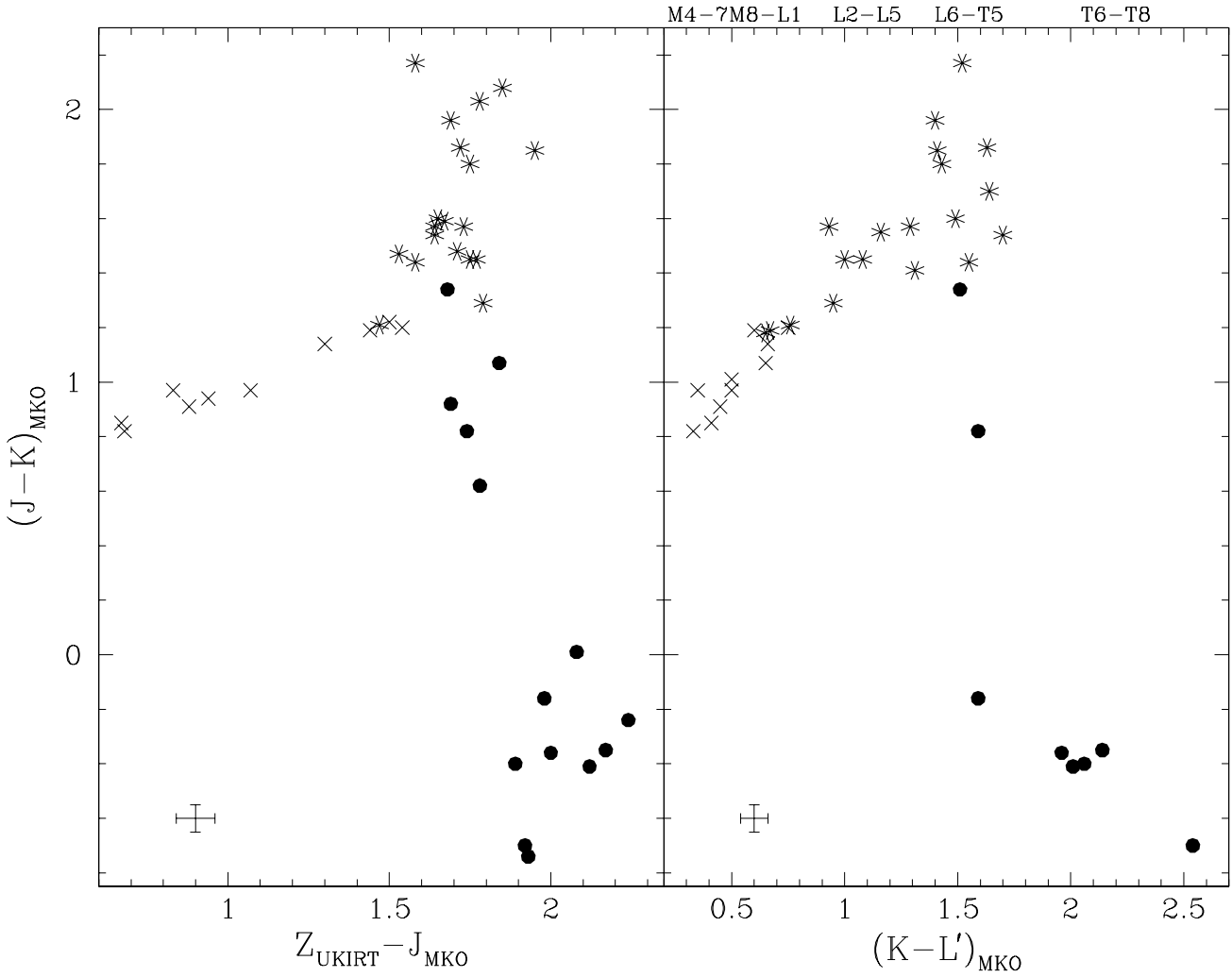}{14truecm}{-90}{70}{70}{-300}{450}
\caption{ Color--color diagrams of $J$--$K$ versus $Z$--$J$ (left) and 
$K$--$L^{\prime}$ (right) for M~dwarfs (crosses), L~dwarfs (asterisks), 
and T~dwarfs (filled circles).
\label{fig6}}
\end{figure}

\clearpage

\begin{deluxetable}{lllrl}
\small
\tablenum{1}
\tablewidth{400pt}
\tablecaption{The Sample}
\tablehead{
\colhead{Name} &  \colhead{R.A.} &  \colhead{Decl.} 
 &\colhead{Distance } & \colhead{Spectral}\nl
 & \multicolumn{2}{c}{(J2000)}& \colhead{Modulus}  & \colhead{Type} \nl  
  & & & \colhead{ $M$--$m$ (error)}  &  \nl 
}
\tablecolumns{5}
\startdata 
BRI 0021--0214  & 00:24:24.6 & $-$01:58:22 &  $-$0.42 (0.08)\tablenotemark{a}   & M9.5  \nl  
2MASSI J0028+15 & 00:28:39.4 & $+$15:01:41 &  \nodata  &  L3  \nl
SDSS 0032+14    & 00:32:59.4 & $+$14:10:37 & \nodata  & L8 \nl   
2MASSW J0036+18 & 00:36:15.9 & $+$18:21:10 & 0.30 (0.02)\tablenotemark{b} & L4  \nl  
SDSS 0107+00    & 01:07:52.3 & $+$00:41:56 & \nodata &  L5.5    \nl  
SDSS 0151+12    & 01:51:41.7 & $+$12:44:30 & \nodata  & T1    \nl   
LHS 11          & 02:00:10.0 & $+$13:03:06 &  1.76 (0.03)\tablenotemark{a}   &  M5  \nl 
DENIS-P J0205--11AB & 02:05:29.0 & $-$11:59:25 &  $-$1.13 (0.10)\tablenotemark{c} & L5.5   \nl 
SDSS 0207+00    & 02:07:42.8 & $+$00:00:56 &  \nodata  &  T4.5 \nl  
SDSS 0236+00    & 02:36:17.9 & $+$00:48:55 & \nodata  & L6.5   \nl
2MASSW J0310+16 & 03:10:59.9 & $+$16:48:16 & \nodata & L9  \nl   
2MASSW J0328+23 & 03:28:42.6 & $+$23:02:05 & \nodata & L9.5   \nl 
2MASP J0345+25  & 03:45:43.2 & $+$25:40:23 &  $-$2.18 (0.04)\tablenotemark{c} & L1    \nl  
SDSS 0423--04   & 04:23:48.6 & $-$04:14:04 & \nodata & T0\nl   
SDSS 0539--00   & 05:39:52.0 & $-$00:59:02 & \nodata&  L5   \nl 
2MASSW J0559--14  &   05:59:19.1 & $-$14:04:48 & \nodata  & T4.5 \nl  
Gl 229B         & 06:10:34.7 & $-$21:51:49 &  1.19 (0.02)\tablenotemark{d} & T6 \nl   
2MASSI J0746+20AB   & 07:46:42.5 & $+$20:00:32  & $-$0.44 (0.02)\tablenotemark{b}  &  L1 \nl   
2MASSI J0825+21 & 08:25:19.6 & $+$21:15:52 & $-$0.48 (0.33)\tablenotemark{e}  & L6  \nl 
SDSS 0830+48    & 08:30:08.1 & $+$48:28:47 & \nodata& L9  \nl    
SDSS 0837--00   & 08:37:17.2 & $-$00:00:18 & \nodata & T0.5 \nl 
2MASSs J0850+10AB  & 08:50:35.9 & $+$10:57:16  &$-$2.47 (0.34)\tablenotemark{c} &  L6\tablenotemark{f}   \nl  
SDSS 0857+57    & 08:57:58.5 & $+$57:08:51 &\nodata& L8\nl  
SDSS 0926+58    & 09:26:15.4 & $+$58:47:21 & \nodata &  T4.5 \nl    
2MASSW J0929+34 & 09:29:33.6 & $+$34:29:52 &\nodata & L8\tablenotemark{f}   \nl  
SDSS 1021--03   & 10:21:09.7 & $-$03:04:20 & \nodata & T3 \nl      
2MASSI J1047+21 & 10:47:53.9 & $+$21:24:23 &  \nodata & T6.5  \nl   
LHS 292         & 10:48:13.0 & $-$11:20:12 &  1.72 (0.04)\tablenotemark{a}   & M6.5   \nl  
LHS 36          & 10:56:28.9 & $+$07:00:52 &  3.11 (0.01)\tablenotemark{a}  &  M6   \nl   
DENIS-P J1058--15   & 10:58:46.5 & $-$15:48:00 & $-$1.22 (0.04)\tablenotemark{c}  &   L3  \nl     
LHS 2347        & 11:05:09.0 & $+$07:06:48 &  \nodata & M5    \nl 
SDSS 1110+01    & 11:10:10.0 & $+$01:16:13 &  \nodata  & T6    \nl  
LHS 2397a       & 11:21:49.0 & $-$13:13:08 &  $-$0.77 (0.07)\tablenotemark{a}   &  M8   \nl    
LHS 315         &  11:47:45.0 & $+$00:48:24  &  2.38 (0.02)\tablenotemark{d}  &   M4   \nl 
2MASSW J1217--03 & 12:17:11.1 & $-$03:11:13 & \nodata  &  T8   \nl    
2MASSW J1225--27 & 12:25:54.3 & $-$27:39:47 & \nodata &  T6  \nl 
DENIS-P J1228--15AB   &  12:28:13.8 & $-$15:47:11 & $-$1.29 (0.11)\tablenotemark{c} & L6   \nl  
LHS 333AB       &  12:33:17.0 & $+$09:01:18 &  1.79 (0.04)\tablenotemark{a} & M4   \nl 
SDSS 1254--01   &  12:54:53.9 & $-$01:22:47 & \nodata  & T2 \nl   
SDSS 1257--01   &  12:57:37.3 & $-$01:13:36 & \nodata   &  L5 \nl      
Kelu-1          &  13:05:40.2 & $-$25:41 06 &  $-$1.40 (0.08)\tablenotemark{c}  &  L3    \nl  
SDSS 1314--00   &  13:14:15.5 & $-$00:08:48 &  \nodata &  L2 \nl     
SDSS 1326--00   &  13:26:29.8 & $-$00:38:32 & \nodata  &  L5.5   \nl
SDSS 1346--00   &  13:46:46.5 & $-$00:31:50 & \nodata &  T6   \nl
GD 165B         &  14:24:39.9 & $+$09:17:16 & $-$2.49 (0.17)\tablenotemark{a} & L3   \nl  
LHS 2924        &  14:28:43.4 & $+$33:10:42 &  $-$0.17 (0.03)\tablenotemark{a} &  M9   \nl    
2MASSW J1439+19 &  14:39:28.4 & $+$19:29:15 &  $-$0.78 (0.02)\tablenotemark{c}  & L1\tablenotemark{f}   \nl   
SDSS 1446+00    &  14:46:00.6 & $+$00:24:52 & \nodata  &  L5   \nl   
LHS 3003        &  14:56:38.0 & $-$28:09:48 & 0.97 (0.04)\tablenotemark{a}  &   M7  \nl    
Gl 570D         &  14:57:15.0 & $-$21:21:51 & 1.14 (0.02)\tablenotemark{d} &  T8  \nl   
TVLM 513--46546  & 15:01:08.3 & $+$22:50:02 &     0.04 (0.11)\tablenotemark{g}   & M8.5 \nl   
2MASSW J1507--16 & 15:07:47.6 & $-$16:27:38 & 0.59 (0.39)\tablenotemark{b}  &  L5\tablenotemark{f} \nl  
2MASSW J1523+30  & 15:23:22.6 & $+$30:14:56 & $-$1.35 (0.05)\tablenotemark{d}  & L8 \nl
SDSS 1624+00    &  16:24:14.4 & $+$00:29:16 & \nodata & T6     \nl  
2MASSW J1632+19  & 16:32:29.1 & $+$19:04:41 & $-$1.20 (0.14)\tablenotemark{c}  & L7.5     \nl  
SDSS 1750+17     & 17:50:33.0 & $+$17:59:04 &  \nodata & T3.5    \nl  
SDSS 2249+00     & 22:49:53.5 & $+$00:44:04 &  \nodata & L5   \nl    
SDSS 2255--00   &  22:55:29.1 & $-$00:34:33 & \nodata & M8.5   \nl  
\nl
\tablecomments{Parallaxes for distance moduli taken from:}
\tablenotetext{a}{\cite{van94}}
\tablenotetext{b}{Dahn 2000 (private communication reported by \cite{k00})}
\tablenotetext{c}{\cite{nofs99}}
\tablenotetext{d}{\cite{hip}}
\tablenotetext{e}{Dahn 1999 (private communication reported by \cite{re01b})}
\tablenotetext{g}{\cite{t95}}
\tablecomments{Spectral types from G02 except:}
\tablenotetext{f}{Spectral type from \cite{k00}}

\enddata

\end{deluxetable}

\newpage

\begin{deluxetable}{rrrl}
\tablenum{2}
\tablecaption{Filter Bandpasses}
\tablehead{
\colhead{Filter} &  \colhead{50\% Cut--On}& \colhead{50\% Cut--Off }
 &  \colhead{Installed}   \nl
\colhead{Name} &  \colhead{Wavelength }& \colhead{Wavelength }
 &  \colhead{in Camera}   \nl
 \colhead{} &  \colhead{$\mu$m}& \colhead{$\mu$m}
 &  \colhead{}   \nl
}
\tablecolumns{4}
\startdata 
UKIRT $Z$ & 0.851 & 1.056 & UFTI \nl
MKO $J$ & 1.165 & 1.329 & UFTI, IRCAM \nl
MKO $H$ & 1.487 & 1.780 & UFTI, IRCAM \nl
MKO $K$ &  2.027 & 2.362 & UFTI, IRCAM \nl
MKO $L^{\prime}$ & 3.429 & 4.107 & IRCAM \nl
MKO $M^{\prime}$ & 4.572 & 4.801 & IRCAM \nl
\nl
\enddata

\end{deluxetable}


\begin{deluxetable}{lrllrl}
\small
\tablenum{3}
\tablecaption{UKIRT UFTI $Z$ Photometry}
\tablehead{
\colhead{Name} &  \colhead{$Z$} &  \colhead{Date} 
& \colhead{Name} &  \colhead{$Z$} &  \colhead{Date} \nl 
\colhead{} &  \colhead{$\pm 0.05$} &  \colhead{(YYYYMMDD)} 
& \colhead{} &  \colhead{$\pm 0.05$} &  \colhead{(YYYYMMDD)} \nl 
\colhead{} &  \colhead{(mag)} &  \colhead{} 
& \colhead{} &  \colhead{(mag)} &  \colhead{} \nl
}
\tablecolumns{6}
\startdata 
BRI 0021--0214 & 13.27  &   20001205  &  2MASSI J1047   &  17.39  &  20001207  
\nl  
2MASSI J0028+15  &  18.50   & 20010114, 20001205  & LHS 292  &  9.76 &  20001204 
\nl 
SDSS 0032+14  &  18.25 & 20001012 &   LHS 36    &  8.11  &  20000303, 20001204  
\nl 
2MASSW J0036+18   &  14.11  & 20001205   &  DENIS-P J1058   & 15.76  &  20001204 
\nl  
SDSS 0107+01    &  17.33   &  20001012   &  LHS 2347 & 13.96   &  20001204, 
19991230   \nl 
SDSS 0151+12  &  18.09   &  20001205   &   LHS 2397a   &  13.13  &  20001204, 
19991230  \nl   
LHS 11   &  8.15  & 20001204  & LHS 315  &   7.17   &  19991230  \nl 
DENIS-P J0205--11AB &  16.01 &  20001204   & 2MASSW J1217  & 17.56  &  20001207  
\nl   
SDSS 0207+00  & 18.71   & 20001205    &  2MASSW J1225 & 16.77 & 20010203 \nl
SDSS 0236+00    &  17.54     &  20001204 & DENIS-P J1228AB   &  16.01 &  
20001204  \nl 
2MASSW J0328+23   &  18.06   &  20001205  &   LHS 333AB  & 7.87  &  19991230 \nl 
SDSS 0423--04    &15.98  &  20001018 &   SDSS 1254  &  16.40 &  20000302 \nl 
SDSS 0539--00   &  15.60   &  20000314 &  Kelu-1  &   15.00  &  20000302   \nl
2MASSW J0559--14  &  15.55   &  20001018  &   SDSS 1346    & 17.73  &  19990620  
\nl 
2MASSI J0746+20AB   & 13.11   &  20001204 &    LHS 2924    &13.35   &  19990824  
\nl
SDSS 0830+48   &16.86  &  20001204   &     Gl 570D   &  16.74    &  20000201 \nl 
SDSS 0837--00   & 18.59 &  20000302 &  2MASSW J1523  &  17.60 &  20000314 \nl
2MASSs J0850+10AB  &  18.15 & 20010114  & SDSS 1624 &   17.32   &  19990620 \nl 
SDSS 0857+57    & 16.52  &  20001204  &   2MASSW J1632  &  17.52   &  20000201  
\nl  
2MASSW J0929+34   &  18.38  &  20001207   &   SDSS 2249   & 18.24&  20001204 \nl  
SDSS 1021--03   &  17.66  &  20000314  &        SDSS 2255   &   17.00 &  
20000923  \nl 
\enddata

\end{deluxetable}

\newpage

\begin{deluxetable}{lrrrrll}
\small
\tablenum{4}
\tablewidth{400pt}
\tablecaption{UKIRT MKO--NIR $JHK$ Photometry}
\tablehead{
\colhead{Name} & \colhead{$J$} & \colhead{$H$} & \colhead{$K$} & \colhead{error} 
& Camera & Date \nl
\colhead{} & \colhead{} & \colhead{} & \colhead{} & \colhead{(mag)} &  & 
(YYYYMMDD) \nl
}
\tablecolumns{7}
\startdata 
2MASSI J0028+15  & 16.65 & 15.56 & 14.57 & 0.05 & UFTI & 20010114 \nl
SDSS 0032+14  &  16.58 & 15.66 & 14.99 & 0.05 & UFTI & 20000921 \nl   
2MASSW J0036+18   &  12.32 & 11.63 & 11.03 & 0.03 & UFTI & 20001205 \nl  
SDSS 0107+00    &   15.75 & 14.56 & 13.58 &   0.03 & UFTI & 19991017, 
20010124\nl  
SDSS 0151+12  &    16.25 &  15.54 & 15.18 & 0.05 & UFTI & 20000921 \nl   
DENIS-P J0205--11AB &  14.43 & 13.61 & 12.99 & 0.05 & IRCAM & 19990919 \nl 
SDSS 0207+00  & 16.63 & 16.66 & 16.62 & 0.05 & UFTI & 20001205 \nl  
SDSS 0236+00    &  16.01 & 15.16 & 14.54 & 0.05 & UFTI & 19991027  \nl
2MASSW J0310+16   & 15.88 & 14.91 & 14.18 & 0.03 & UFTI & 20001205 \nl   
2MASSW J0328+23   &  16.35 & 15.47 & 14.87 & 0.03 & UFTI & 20001205   \nl 
SDSS 0423--04    & 14.30  & 13.51&  12.96 & 0.03 & UFTI & 20001013 \nl   
2MASSW J0559--14  & 13.57 & 13.64 & 13.73 &   0.03 & UFTI & 20001013 \nl  
2MASSI J0746+20AB   &  11.64 & 11.01 & 10.43 & 0.03 & IRCAM & 20001206 \nl   
2MASSI J0825+21 &  14.89 & 13.81 & 12.93 & 0.03 & UFTI &   20010123   \nl 
SDSS 0830+48   & 15.22 & 14.40 &  13.68 & 0.03 & IRCAM & 20001119, 20001206 \nl    
2MASSs J0850+10AB  &  16.20 & 15.21 & 14.35 &  0.03 & UFTI & 20001207 \nl  
SDSS 0857+57    & 14.80 & 13.80 &  12.94 & 0.03 & IRCAM &  20001206 \nl  
SDSS 0926+58   &  15.47 & 15.42 & 15.50 & 0.03 & UFTI &   20010123  \nl    
2MASSW J0929+34   &  16.69 & 15.62 & 14.73 & 0.03 & UFTI & 20001207 \nl  
2MASSI J1047+21   &   15.46 & 15.83 & 16.20 & 0.03 & UFTI & 20001207 \nl   
SDSS 1110+01    &  16.12 & 16.22 & 16.05  & 0.05 & UFTI & 20010218 \nl  
2MASSW J1217--03  &  15.56 & 15.98 & 15.92 & 0.03 & UFTI & 20001207  \nl    
2MASSW J1225--27 & 14.88 & 15.17 & 15.28 & 0.03 & UFTI & 20001207 \nl 
SDSS 1257--01  &   15.64 & 14.68 & 14.06 & 0.03 & UFTI & 20000314 \nl      
SDSS 1314--00   &  16.33 & 15.84 & 15.28 & 0.05 & UFTI & 20000314  \nl     
SDSS 1326--00 &  16.19 & 15.06 & 14.13 & 0.05 & UFTI & 20000314, 20010123\nl
2MASSW J1439+19  &  12.66 & 12.05 & 11.47 &  0.03 & UFTI &   20010123  \nl   
SDSS 1446+00  &  15.56 & 14.59 & 13.80 & 0.05 & UFTI & 20000314  \nl   
2MASSW J1507--16 &  12.70 & 11.90 & 11.29 &   0.03 & UFTI &   20010123 \nl  
2MASSW J1523+30  & 15.95 & 15.05 & 14.35 & 0.05 & UFTI & 20000314  \nl
2MASSW J1632+19  &  15.77 & 14.68 & 13.97 & 0.05 & UFTI & 20000201    \nl  
SDSS 1750+17   &   16.14 & 15.94 & 16.02 &  0.05 & UFTI & 20010218  \nl  
SDSS 2249+00   &   16.46 & 15.42 & 14.43 & 0.05 & UFTI & 20010116 \nl    
SDSS 2255--00   &   15.50 &  14.80 & 14.28  & 0.05 & UFTI & 20000921 \nl  

\enddata

\end{deluxetable}

\newpage

\begin{deluxetable}{lrlrl}
\small
\tablenum{5}
\tablecaption{UKIRT IRCAM $L^{\prime}$ and $M^{\prime}$ Photometry}
\tablehead{
\colhead{Name} &  \colhead{$L^{\prime}$ (error)} &  \colhead{Date} & 
\colhead{$M^{\prime}$ (error)} &  \colhead{Date} \nl
\colhead{} &  \colhead{(mag)} &  \colhead{(YYYYMMDD)} & \colhead{(mag)} &  
\colhead{(YYYYMMDD)} \nl
}
\tablecolumns{5}
\startdata 
2MASSW J0036+18   &  10.08 (0.05) & 20010120 & 10.35 (0.05) & 20010121, 20010122 
 \nl 
SDSS 0107+00    &  12.06 (0.07)  & 20001120 & \nodata & \nodata\nl  
2MASSW J0310+16   &  12.54 (0.05)  & 20010218 & \nodata &\nodata\nl   
SDSS 0423--04    & 11.45 (0.05) & 20001120 & \nodata &\nodata \nl   
SDSS 0539+00   &   11.32 (0.05) & 20001206 & \nodata &\nodata \nl 
2MASSW J0559--14  & 12.14 (0.05) & 20001120 & 12.15 (0.15) & 20010121, 20010122  
\nl  
2MASSI J0746+20AB   &  9.67 (0.03) & 20001206 & \nodata & \nodata \nl   
SDSS 0830+48   &  11.98 (0.05) & 20001206 & \nodata & \nodata \nl    
2MASSI J0825+21 & 11.53 (0.03) & 20010523 & \nodata& \nodata\nl 
2MASSs J0850+10AB  &  12.94 (0.05)  & 20010120 & \nodata& \nodata  \nl  
SDSS 0857+57    & 11.31 (0.05) & 20001206 & 11.50 (0.10) & 20010121\nl  
LHS 2397a   & 10.03 (0.02) & 20010120 &  \nodata&  \nodata \nl    
2MASSW J1217--03  &  13.96 (0.05) & 20010523 & \nodata&  \nodata \nl 
2MASSW J1225--27 &   13.22 (0.08) & 20010120 & & \nl 
SDSS 1254--01  & 12.25 (0.05) & 20000303 & 12.65 (0.20) & 20010121, 20010122 \nl   
LHS 2924    & 10.12 (0.03) & 20000303, 20010120  & \nodata  & \nodata \nl    
2MASSW J1439+19  &  10.80 (0.05) & 20010120 &   \nodata  &  \nodata \nl   
2MASSW J1507--16 &  9.98 (0.03) & 20010120 &   \nodata &  \nodata \nl  
2MASSW J1523+30  & 12.86 (0.05)  & 20010523 & \nodata&  \nodata \nl 
SDSS 1624+00 &  13.60 (0.04) & 19990920, 20010523 &    \nodata  & \nodata \nl  
2MASSW J1632+19  &  12.54 (0.05) & 20000515  &  \nodata &  \nodata \nl  
\enddata

\end{deluxetable}

\newpage

\begin{deluxetable}{llrrrrrrrrr}
\small
\tablenum{6}
\tablecaption{Infrared Magnitudes and Colors, Sorted by Spectral Type}
\tablehead{
\colhead{Name} &  \colhead{Type} &  \colhead{$M_K$} 
& \colhead{$Z$--$J$} & \colhead{$J$}& 
\colhead{$J$--$K$} &\colhead{$J$--$H$} &\colhead{$H$--$K$} &
\colhead{$K$}& \colhead{$K$--$L^{\prime}$} & \colhead{$L^{\prime}$--$M^{\prime}$}\nl
}
\tablecolumns{10}
\startdata 
LHS 315\tablenotemark{a}  &    M4   &  8.03  &   0.67 &  6.50 &  0.85 &0.51 & 
0.34 & 5.65 & 0.41 & \nodata \nl 
LHS 333AB\tablenotemark{a}  &   M4  & 7.87  &   0.88  & 6.99 &  0.91 &0.50 & 
0.41 & 6.08 &  0.45 & \nodata \nl 
LHS 11\tablenotemark{a}   &  M5  &  8.41   &  0.68 &  7.47  &  0.82 &0.48 & 0.35 
& 6.65 &  0.33& \nodata \nl 
LHS 2347 &  M5   &  \nodata   & 0.94  & 13.02 &  0.94  &  0.52  &  0.42 &  12.08 
&  \nodata & \nodata\nl 
LHS 36\tablenotemark{b}    &   M6   &  9.17  &   1.07 &  7.03   & 0.97   & 0.55 
& 0.43 & 6.06 &  0.35 & \nodata\nl  
LHS 292\tablenotemark{b}  & M6.5 & 9.67    &   0.83  &  8.92 &  0.97  & 0.53 & 
0.44 & 7.95  & 0.50& \nodata \nl 
LHS 3003\tablenotemark{b}   &   M7  & 9.90  & \nodata &  9.94 &  1.01 & 0.51 & 
0.50 & 8.93 &  0.50& \nodata \nl   
LHS 2397a   &  M8 & 9.92   & 1.30 &  11.83  & 1.14  &  0.57  &  0.57 &  10.69  & 
0.66 & \nodata \nl   
SDSS 2255--00   &  M8.5   & \nodata & 1.50 &  15.50  & 1.22  &  0.70  &  0.52  & 
14.28  & \nodata& \nodata\nl  
TVLM 513--46546\tablenotemark{b}  &   M8.5 &  10.73    & \nodata &  11.76 & 1.07 
& 0.58 &   0.49 &  10.69 &  0.65& \nodata\nl 
LHS 2924    &  M9  & 10.55  &  1.44  &  11.91  & 1.19  &  0.64  &  0.55 &  10.72 
 & 0.60& \nodata \nl 
BRI 0021--0214\tablenotemark{b}  &   M9.5 &  10.11 &   1.54  & 11.73  & 1.20 & 
0.63 & 0.57 &  10.53 &  0.75& \nodata \nl
2MASSI J0746+20AB   &  L1 &  9.99 &  1.47 &  11.64 &  1.21  &  0.63 &0.58 &  
10.43 &  0.76& \nodata\nl
2MASP J0345+25\tablenotemark{b}  &  L1 &  10.48  & \nodata  & 13.84  & 1.18 & 
0.63 &0.54  & 12.66 &  0.65& \nodata\nl  
2MASSW J1439+19\tablenotemark{c}  &  L1 &  10.69    &  \nodata  &  12.66 & 1.19 
& 0.61 &   0.58 &  11.47 &  0.67& \nodata\nl   
SDSS 1314--00   &   L2 &  \nodata  & \nodata&  16.33 &  1.05  &  0.49  &0.56 &  
15.28 & \nodata& \nodata\nl   
2MASSI J0028+15  &   L3 &  \nodata   &  1.85  & 16.65  &  2.08  &  1.09  &  0.99 
&  14.57 & \nodata & \nodata\nl
DENIS-P J1058--15\tablenotemark{b}   &  L3  & 11.33  &  1.64 &   14.12 & 1.57 & 
0.84 & 0.74 &  12.55 &  0.93& \nodata\nl 
GD 165B\tablenotemark{b}    &  L3 & 11.60  & \nodata &  15.64 &  1.55  & 0.89 & 
0.66 &14.09 &  1.16& \nodata \nl 
Kelu-1\tablenotemark{b}  &    L3  &  10.38  &  1.77  &  13.23  & 1.45  & 0.78 & 
0.67 & 11.78 &  1.00& \nodata\nl  
2MASSW J0036+18   & L4   & 11.33  &  1.79  &  12.31 &  1.29  &  0.69  &0.60 &  
11.03 &  0.95& --0.27 \nl 
2MASSW J1507--16\tablenotemark{c} &  L5  &  11.88     & \nodata  & 12.70 & 1.41 
& 0.80 &  0.61  & 11.29  & 1.31& \nodata\nl 
SDSS 0539--00   &  L5 & \nodata&  1.75  &  13.85  & 1.45  &  0.81   & 0.64  & 
12.40  & 1.08 & \nodata\nl 
SDSS 1257--01  & L5 & \nodata   &  \nodata  & 15.64  & 1.58  &  0.96 &   0.62 & 
14.06  &  \nodata& \nodata\nl 
SDSS 1446+00  & L5  & \nodata    & \nodata  & 15.56  &  1.76  &  0.97  &  0.79 & 
 13.80 & \nodata & \nodata\nl 
SDSS 2249+00   &L5  &  \nodata & 1.78  & 16.46 &  2.03   & 1.04 &   0.99 &14.43 
&  \nodata & \nodata\nl  
DENIS-P J0205--11AB &  L5.5  & 11.86  &  1.58  &  14.43  & 1.44  &  0.82  &0.62  & 
12.99 &  1.55& \nodata\nl 
SDSS 0107+00    &  L5.5   &\nodata  & 1.58  &  15.75 &  2.17  &  1.19   &0.98  & 
13.58 &  1.52& \nodata \nl 
SDSS 1326--00 &  L5.5   & \nodata    & \nodata  & 16.19 &  2.06  &  1.13& 0.93  
& 14.13  & \nodata& \nodata\nl
2MASSI J0825+21 & L6   & 12.45    &  \nodata  & 14.89  & 1.96  &  1.08  &  
0.88  & 12.93 & 1.40 & \nodata\nl 
2MASSs J0850+10AB\tablenotemark{c}  & L6   & 11.88   & 1.95 & 16.20&1.85 & 0.99& 
0.86&  14.35 & 1.41 & \nodata \nl 
DENIS-P J1228--15AB\tablenotemark{b}   & L6 & 11.42    &  1.73  &  14.28 & 
1.57 & 0.88&  0.69  & 12.71 &  1.29& \nodata \nl  
SDSS 0236+00    &  L6.5  & \nodata  &  1.53  & 16.01 &  1.47  &  0.85  &0.62  & 
14.54 &  \nodata& \nodata\nl
2MASSW J1632+19  &  L7.5 & 12.77   &  1.75  & 15.77 &  1.80  &  1.09  &0.71 &  
13.97 &  1.43& \nodata\nl  
2MASSW J0929+34\tablenotemark{c}   &  L8  &\nodata  & 1.69 & 16.69 &  1.96& 1.07 
& 0.89&  14.73&  \nodata & \nodata\nl  
2MASSW J1523+30  &L8  & 13.00   &  1.65 &  15.95 &   1.60 &   0.90 &   0.70  & 
14.35 & 1.49 & \nodata\nl
SDSS 0032+14  &  L8 & \nodata   & 1.67  &  16.58  & 1.59  &  0.92  &0.67 &  
14.99 &  \nodata & \nodata\nl
SDSS 0857+57    &L8 &\nodata& 1.72&  14.80 & 1.86 &  1.00 &  0.86&  12.94 & 
1.63& --0.19\nl 
SDSS 0830+48   & L9  & \nodata& 1.64 &  15.22 &  1.54  &  0.82 &   0.72 &  13.68 
&  1.70& \nodata\nl   
2MASSW J0310+16   & L9  & \nodata   & \nodata  & 15.88  & 1.70  &  0.97&  0.73 
 & 14.18 &  1.64& \nodata\nl 
 2MASSW J0328+23   & L9.5  & \nodata   &  1.71  & 16.35 &  1.48 &   0.88 & 0.60 
&  14.87 & \nodata& \nodata\nl 
SDSS 0423--04    & T0 & \nodata & 1.68 &  14.30 &  1.34  &  0.79  &  0.55 &  
12.96 &  1.51& \nodata\nl   
SDSS 0837--00   & T0.5 & \nodata  & 1.69 & 16.90 & 0.92 & 0.69 &  0.23 & 15.98 & 
\nodata & \nodata\nl 
SDSS 0151+12  & T1 & \nodata  &  1.84   & 16.25 &  1.07  &  0.71  &  0.36  & 
15.18 & \nodata & \nodata \nl   
SDSS 1254--01  & T2 & \nodata     & 1.74  & 14.66 &  0.82  &  0.53  &  0.29  & 
13.84  & 1.59 & --0.40 \nl 
SDSS 1021--03   & T3  & \nodata  & 1.78  & 15.88 &  0.62  &  0.47  &  0.15  & 
15.26 & \nodata& \nodata \nl
SDSS 1750+17   & T3.5 &  \nodata &   \nodata & 16.14 &  0.12  &  0.20 &  --0.08  & 
16.02   & \nodata& \nodata  \nl  
2MASSW J0559--14  &  T4.5  & \nodata   & 1.98 &  13.57 &  --0.16 &  --0.07 &  --0.09 
&  13.73 &  1.59 & --0.01\nl 
SDSS 0207+00  &  T4.5 & \nodata    & 2.08  & 16.63  & 0.01 &  --0.03 & 0.04 & 
16.62  & \nodata & \nodata\nl  
SDSS 0926+58   & T4.5   & \nodata &  \nodata & 15.47&  --0.03 &  0.05 & --0.08&  
15.50 & \nodata& \nodata \nl    
2MASSW J1225--27 &   T6   & \nodata &  1.89 &  14.88 &--0.40 & 
--0.29 & --0.11 &  15.28  & 2.06 & \nodata\nl 
Gl 229B\tablenotemark{d}   &  T6   & 15.55  &  2.17  & 14.01 &  --0.35 &--0.35 & 
0.00 & 14.36 &  2.14 & \nodata\nl   
SDSS 1110+01   &  T6  &\nodata   &  \nodata & 16.12  &0.07 & 
--0.10 & 0.17  & 16.05   & \nodata  & \nodata\nl 
SDSS 1346--00\tablenotemark{b}    &  T6   & \nodata &    2.24  & 15.49  &--0.24 & 
--0.35 &  0.11 &  15.73  & \nodata& \nodata\nl
SDSS 1624+00\tablenotemark{b} &   T6  & \nodata   & 2.12  & 15.20 &--0.41 & 
--0.28 & --0.13  & 15.61 &  2.01 & \nodata\nl  
2MASSI J1047+21   &  T6.5  &  \nodata &  1.93  & 15.46  &--0.74 & 
--0.37 & --0.37 &  16.20 &  \nodata& \nodata \nl 
2MASSW J1217--03  &  T8   & \nodata &  2.00 &  15.56  &--0.36 & 
--0.42 & 0.06 &  15.92 & 1.96 & \nodata\nl 
Gl 570D   &   T8   & 16.66    & 1.92  & 14.82  & --0.70  & --0.46 &  --0.24  & 
15.52 &  2.54& \nodata\nl   
\nl
\tablenotetext{a}{$JHK$ on MKO system derived from transformations given in 
\cite{haw01}}
\tablenotetext{b}{$JHK$ magnitudes synthesized from flux calibrated spectrum}
\tablenotetext{c}{L spectral type taken from \cite{k00}}
\tablenotetext{d}{$ZJHKL^{\prime}$ magnitudes synthesized from flux calibrated
spectrum}
\tablecomments{Uncertainty in spectral type is $\sim$0.5 subclasses; 
error in $M_K$ is $\sim$10\% except for objects with more uncertain
parallaxes, see Table 1; error in colors is typically $\sim$5\% except for 
$L^{\prime}$--$M^{\prime}$, see Table 5.}
\enddata

\end{deluxetable}

\newpage

\begin{deluxetable}{llrrr}
\tablenum{7}
\tablecaption{Luminosity and Temperature}
\tablehead{
\colhead{Name} &  \colhead{Type}& \colhead{BC$_K$ }
 &  \colhead{log$_{10}~L/L_{\odot}$} 
 & \colhead{$T_{\rm eff}$ (K)}  \nl
}
\tablecolumns{5}
\startdata 
LHS 36    &   M6   &   2.97 & $-$2.97 &  2650---2850   \nl  
LHS 292  & M6.5 & 2.95  &  $-$3.16 &  2500---2700 \nl 
LHS 3003   &   M7  &  3.03&  $-$3.28 & 2400---2650  \nl   
TVLM 513--46546  &   M8.5 &   3.13  & $-$3.65 & 2000---2250  \nl 
BRI 0021--0214  &   M9.5 &    3.17 & $-$3.43  & 2200---2450  \nl
2MASP J0345+25  &  L1 &    3.20 & $-$3.59 & 2050---2350 \nl  
2MASSI J0746+20AB   &  L1 &   3.26 & $-$3.40 & 1900---2200\tablenotemark{*}\nl
Kelu-1  &    L3  &   3.27 &  $-$3.57 & 2100---2350 \nl  
DENIS-P J1058--15   &  L3  &   3.31 & $-$3.98  & 1700---1950 \nl 
GD 165B    &  L3 &   3.27& $-$4.06 & 1650---1850 \nl 
2MASSW J0036+18 & L4 & 3.34 & $-$3.97 & 1700---1950 \nl 
DENIS-P J0205--11AB &  L5.5  &  3.18  & $-$4.12 & 1400---1600\tablenotemark{*} \nl 
DENIS-P J1228--15AB   & L6 &  3.27  & $-$4.00 & 1450---1650\tablenotemark{*} \nl  
2MASSI J0825+21  &  L6  &  3.31  & $-$4.40 & 1400---1600 \nl
2MASSW J1632+19  &  L7.5 &    3.19 & $-$4.48 & 1350---1550 \nl 
2MASSW J1523+30  &L8  &  3.17 &  $-$4.57 & 1250---1500 \nl
Gl 229B   &  T6   &  2.22 & $-$5.21 & 870---1030\nl   
Gl 570D   &   T8   &  1.90& $-$5.53 & 784---824 \nl   
\nl
\tablenotetext{*}{
Assuming equal contribution to luminosity from each component}

\enddata

\end{deluxetable}


\begin{deluxetable}{lrrrrlrrrr}
\small
\tablenum{8}
\tablecaption{$K-L^{\prime}$, $K-M^{\prime}$: Model Comparison }
\tablehead{
\colhead{Type} &  \colhead{$\sim T _{\rm eff}$}& \multicolumn{3}{c}{$K-L{^\prime}$ } &
\colhead{Type} &  \colhead{$\sim T _{\rm eff}$}& \multicolumn{3}{c}{$K-M^{\prime}$ } \nl
\colhead{} &  \colhead{K}& \colhead{Observed} &  \multicolumn{2}{c}{Calculated}  &
\colhead{} &  \colhead{K}&  \colhead{Observed} &  \multicolumn{2}{c}{Calculated}  \nl
 \colhead{} &  \colhead{}&  \colhead{} &  \colhead{Dusty\tablenotemark{a}}
 &  \colhead{Settled\tablenotemark{b}} &
\colhead{} &  \colhead{}&  \colhead{} &  \colhead{Dusty\tablenotemark{a}}
 &  \colhead{Settled\tablenotemark{b}} \nl
}
\tablecolumns{10}
\startdata 
L1 & 2100 & 0.7 & 1.0   & \nodata  &  L4   &   1800        & 0.7$\pm$0.1 &  1.0 & \nodata  \nl
L8 & 1400 & 1.6 & 1.8   & \nodata  &  L8   &   1400        & 1.4$\pm$0.1 &  2.0 & \nodata  \nl
T0 & 1300 & 1.5 & 2.1   & \nodata  &  T2   &   1300---1000 & 1.2$\pm$0.2 &  2.1---3.3 & \nodata \nl 
T6 &  950 & 2.0 & 3.3   &     2.3  &  T4.5 &   1300---1000 & 1.6$\pm$0.2 &  2.1---3.3 &  $\sim$3.0\nl
\nl
\tablenotetext{a}{Chabrier et al. (2000) for ages 
0.1---10~Gyr, corresponding to log$g\approx$4.2---5.4}
\tablenotetext{a}{Burrows et al. (1997) for  log$g=$4.5---5.0 
corresponding to ages $\approx$0.3---1~Gyr}

\enddata

\end{deluxetable}

\clearpage


\begin{thebibliography}{}

\bibitem[Ackerman \& Marley 2001]{ack01} Ackerman, A.~S., \& Marley, M.~S. 2001, 
\apj, 556, 872
\bibitem[Allard et al. 2001]{all01} Allard, F., Hauschildt, P.~H., Alexander, D.~R., 
Tamanai, A. \& Schweitzer, A. 2001, \apj, 556, 357
\bibitem[Bailer--Jones \& Mundt 2001]{bai01}Bailer--Jones, C.A.L. \& Mundt, R. 
2001, \aap, 367, 218
\bibitem[Becklin \& Zuckerman 1988]{bz88} Becklin, E.~E., \& Zuckerman, B. 1988, 
\nat,  336, 656
\bibitem[Beichman et al. 1998]{2m}Beichman, C.A., Chester, T.J., Skrutskie, M., 
Low, F.J. \& Gillett, F. 1998, \pasp, 110, 480
\bibitem[Bessell \& Brett 1988]{bb88}Bessell, M.S., \& Brett, J.M. 1988, \pasp, 
100, 1134
\bibitem[Borysow, Jorgensen \& Zheng 1997]{bor97} Borysow, A., Jorgensen, U.~G., 
Zheng, C. 1997, \aap, 324, 185
\bibitem[Burgasser et al. 1999]{bur99}Burgasser, A.~J., et al. 1999, \apj, 522, 
L65
\bibitem[Burrows et al. 1997]{bur97}Burrows, A., Marley, M., Hubbard, W.~B., 
Lunine, J.~I., Guillot, T., Saumon, D., Freedman, R., Sudarsky, D., \& Sharp, C.
1997, \apj, 491, 856
\bibitem[Burrows, Marley, \& Sharp 2000]{bur00}Burrows, A., Marley, M.~S., \& 
Sharp, C.~M. 2000, \apj, 531, 438
\bibitem[Burrows et al. 2001]{burrows01}Burrows, A., Hubbard, W.~B., Lunine, J.~I.
\& Liebert, J. 2001, Reviews of Modern Physics (astro-ph/0105004)
\bibitem[Chabrier et al. 2000]{cbah00}Chabrier, G., Baraffe, I., 
Allard, F., Hauschildt, P.~H. 2000, \apj, 542, 464
\bibitem[Cuby et al. 1999]{cub99} Cuby, J. G., Saracco, P., Moorwood, A.~F.~M., 
S'Odorico. S.,
Lidman, C., Comer\'{o}n, F., \& Spyromillo, J. 1999, \aap, 349, L41
\bibitem[Dahn et al. 2000]{nofs99} Dahn, C. et al. 2000, Giant Planets to
Cool Stars, ASP Conf. Ser., Edited by C. Griffith and M. Marley 
\bibitem[Delfosse et al. 1997]{d97} Delfosse, X., Tinney, C.~G., Forveille, T.,
Epchtein, N., Bertin, E., Borsenberger, J., Copet, E., De Batz, B., Fouque, P.,
Kimeswenger, S., Le Bertre, T., Lacombe, F., Rouan, D., Tiphene, D. 1997, 
\aap,  327, L25
\bibitem[Fan et al. 2000]{fan00}Fan, X., et al.\ 2000, \aj, 119, 928
\bibitem[Fukugita et al. 1996]{fuk96}Fukugita, M., Ichikawa, T., Gunn, J.E., 
Doi, M., Shimasaku, K. \& Schneider, D.P. 1996, \aj, 111, 1748
\bibitem[Fegley \& Lodders 1996]{fl96}Fegley, B., \& Lodders, K.\ 1996, \apj, 
472, L37
\bibitem[Geballe et al. 1996]{g96}Geballe, T.~R., Kulkarni, S.~R., Woodward, 
C.~E.; Sloan, G.~C. 1996, \apj, 467, L101
\bibitem[Geballe et al. 2001]{g01b} Geballe, T.~R., Saumon, D., Leggett, 
S.~K., Knapp, G.~R., Marley, M.~S., \& Lodders, K. 2001, \apj, 556, 373
\bibitem[Geballe et al. 2002]{g01a}Geballe, T.~R., et al. 2001a, \apj, submitted 
(G02) 
\bibitem[Hawarden et al. 2001]{haw01}Hawarden, T.~G., Leggett, S.~K., Letawsky, 
M.~B., Ballantyne, D.~R., \& Casali, M.~M. 2001, \mnras, 325, 563
(See also 
http://www.jach.hawaii.edu/JACpublic/UKIRT/astronomy/calib/fs\_newJHK.html)
\bibitem[Kirkpatrick, Beichman, \& Skrutskie 1997]{k97} Kirkpatrick, J.~D.,
Beichman, C.~A., \& Skrutskie, M.~F. 1997, \apj, 476, 311
\bibitem[Jones et al. 1996]{jon96}Jones, H.R.A., Longmore, A.J., Allard, F.,
 Hauschildt, P.H., 1996, \mnras, 280, 77
\bibitem[Kirkpatrick et al. 1999a]{k99a} Kirkpatrick, J.~D., Reid, I.~R., 
Liebert, J., Cutri, R.~M., Nelson, B., Beichman, C.~A., Dahn, C.~C., Monet, D.~A., 
Gizis, J.~E., \& Skrutskie, M. F. 1999a, \apj, 519, 802
\bibitem[Kirkpatrick et al. 1999b]{k99b} Kirkpatrick, J.~D., Allard, F., Bida, 
T., Zuckerman, B., Becklin, E.~E., Chabrier, G., Baraffe, I. 1999b, \apj, 519, 834
\bibitem[Kirkpatrick et al. 2000]{k00}Kirkpatrick, J.~D., et al.  2000, \aj, 
120, 447
\bibitem[Krisciunas, Margon, \& Szkody 1998]{kr98} Krisciunas, K., Margon, B., 
\& Szkody, P. 1998, \pasp, 110, 1342
\bibitem[Leggett 1992]{l92}Leggett, S.K., 1992, \apjs,  82, 351
\bibitem[Leggett et al. 1998]{l98}Leggett, S.K., Allard, F.,
\& Hauschildt, P.H., 1998, \apj,  509, 836
\bibitem[Leggett et al. 1999]{l99}Leggett, S.~K., Toomey, D.~W., Geballe, 
T.~R., \& Brown, R.~H. 1999, \apj, 517, L139
\bibitem[Leggett et al. 2000a]{l00a}Leggett, S.~K., Allard, F.,  Dahn,
C., Hauschildt, P.~H., Kerr, T.~H., \& Rayner, J.\ 2000a, \apj,  535, 965 
\bibitem[Leggett et al. 2000b]{l00b}Leggett, S.~K.,  et al. 2000b, \apjl, 536, 
L35
\bibitem[Leggett et al. 2001]{l01}Leggett, S.~K., Allard, F.,  Geballe, T.~R.,
Hauschildt, P.~H., \& Schweitzer, A.\ 2001, \apj, 548, 908
\bibitem[Liebert et al. 2000]{lie00}Liebert, J., Reid, I.~N., Burrows, A., 
Burgasser, A.~J., Kirkpatrick, J.~D., \& Gizis, J.~E.\ 2000, \apj, 533, L155
\bibitem[Lodders 1999]{lo99} Lodders, K. 1999, \apj, 519, 793
\bibitem[Mart\'{\i}n, Brandner, \& Basri 1999a]{mar99a} Mart\'{\i}n, E.L., 
Brandner, W., \& Basri, G. 1999a, Science, 283, 1718 
\bibitem[Mart\'{\i}n et al. 1999b]{mar99b}Mart\'{\i}n, E.L., Delfosse, X., 
Basri, G., Goldman, B., Forveille, T., \& Zapatero Osorio, M.~R. 1999b, \aj, 118, 2466
\bibitem[Matthews et al. 1996]{mat96}Matthews, K., Nakajima, T., Kulkarni,
S.~R., \& Oppenheimer, B.~R. 1996, \aj,  112, 1678
\bibitem[Nakajima et al. 1995]{nak95} Nakajima, T., Oppenheimer, B.~R.,
Kulkarni, S.~R., Golimowski, D.~A., Matthews, K., \& Durrance, S.~T. 1995, \nat, 
 378, 463
\bibitem[Noll, Geballe, \& Marley 1997]{noll97}Noll, K.~S., Geballe, T.~R., 
\& Marley, M.~S. 1997, \apjl, 489, L87
\bibitem[Noll et al. 2000]{noll00}Noll, K.~S., Geballe, T.~R., Leggett, S.~K.,
\& Marley, M.~S. 2000, \apjl, 541, L75
\bibitem[Oppenheimer et al. 1998]{opp98}Oppenheimer, B.~R., Kulkarni, S.~R., 
Matthews, K. \& van Kerkwijk, M.~H. 1998, \apj, 502, 932 
\bibitem[Perryman et al. 1997]{hip}Perryman, M.~A.~C., et al. 1997, \aap, 323, 
L49
\bibitem[Reid et al. 2001a]{re01a}Reid, I.~N., Burgasser, A.~J., Cruz, K.~L., 
Kirkpatrick, J.~D. \& Gizis, J.~E. 2001a, \aj, 121, 1710
\bibitem[Reid et al. 2001b]{re01b}Reid, I.~N., Gizis, J.~E., Kirkpatrick, 
J.~D., \& Koerner, D.~W. 2001b, \aj, 121, 489
\bibitem[Ruiz, Leggett, \& Allard 1997]{rla97} Ruiz,  M.~T., Leggett, S.~K., 
\& Allard, F. 1997, \apj,  491, L107
\bibitem[Saumon et al. 2000]{sa00} Saumon, D., Geballe, T.~R., Leggett, S.~K., 
Marley, M.~S., Freedman, R.~S., Lodders, K., Fegley, B., \& Sengupta, S.~K.
2000, \apj, 541, 374
\bibitem[Simons \& Tokunaga 2002]{si01}Simons, D., \& Tokunaga, A. 2001,
\pasp, submitted
\bibitem[Stephens et al. 2001]{ds01}Stephens, D.~S., Marley, M.~S., Noll, K.~S. 
\& Chanover, N., 2001, \apj, 556, L97
\bibitem[Strauss et al. 1999]{str99}Strauss, M.~A., et al. 1999, \apj, 522, L61
\bibitem[Tinney et al. 1995]{t95}Tinney, C.~G., Reid, I.~N., Gizis, J., \& 
Mould, J.~R. 1995, \aj,  110, 3014
\bibitem[Tokunaga \& Simons 2002]{to01}Tokunaga, A., \& Simons, D. 2002, \pasp, 
submitted
\bibitem[Tsuji et al. 1996]{tsu96}Tsuji, T., Ohnaka, K., Aoki, W., \& Nakajima, 
T. 1996, \aap, 308, L29
\bibitem[Tsuji, Ohnaka, \& Aoki 1999]{tsu99}Tsuji, T., Ohnaka, K., \& Aoki, W.
1999, \apj, 520, L119
\bibitem[Tsvetanov et al. 2000]{tsv00}Tsvetanov, Z. I., et al. 2000, \apj, 531, 
L61
\bibitem[van Altena, Lee, \& Hoffleit 1994]{van94} van Altena, W.~F., Lee, 
J.~T., \& Hoffleit, E.~D. 1994, The General Catalogue of Trigonometric 
Parallaxes (New Haven: Yale University Observatory)
\bibitem[York et al. 2000]{sdss}York, D.G., et al. 2000, \aj, 120, 1579

\end{thebibliography}
\end{document}